\documentclass[%
 a4paper,
 reprint,
 superscriptaddress,
 nofootinbib,
 amsmath,
 amssymb,
 aps,
prd,
floatfix,
]{revtex4-1}

\usepackage[english]{babel}
\usepackage[utf8]{inputenc}
\usepackage[T1]{fontenc}

\usepackage[colorinlistoftodos, color=green!40, prependcaption]{todonotes}

\usepackage{graphicx}

\usepackage{booktabs}

\usepackage[pdftitle={Article}, pdfauthor={Author}]{hyperref}

\usepackage{journals}

\usepackage{dcolumn}

\usepackage{marvosym}
\usepackage{svrsymbols}
\usepackage[export]{adjustbox}
\usepackage[capitalize]{cleveref}

\usepackage{bm}

\usepackage{enumitem}

\usepackage{makebox}

\usepackage{soul}

\newcommand{\ie}{i.e.,~}

\newcommand{\etal}{{\it et al.}}

\newcommand{\unit}[1]{%
    \ensuremath{\, \mathrm{#1}}}

\newcolumntype{d}[1]{D{.}{.}{#1}}
\newcommand{\mc}[1]{\multicolumn{1}{c}{#1}}

\newcommand{\mcA}{\mathcal{A}}
\newcommand{\mcB}{\mathcal{B}}
\newcommand{\mcH}{\mathcal{H}}
\newcommand{\mcK}{\mathcal{K}}
\newcommand{\mcL}{\mathcal{L}}
\newcommand{\mcM}{\mathcal{M}}
\newcommand{\mcP}{\mathcal{P}}
\newcommand{\mcQ}{\mathcal{Q}}
\newcommand{\mcW}{\mathcal{W}}
\newcommand{\mcX}{\mathcal{X}}
\newcommand{\mcY}{\mathcal{Y}}

\newcommand{\rst}{r\sin\theta}
\newcommand{\rrst}{r^2\sin\theta}
\newcommand{\rrsst}{r^2\sin^2\theta}

\begin{document}

\title{Dynamics of Fast Rotating Neutron Stars: An Approach in the Hilbert Gauge}

\author{Christian J. Kr\"uger}
    \email{christian.krueger@tat.uni-tuebingen.de}
    \affiliation{Theoretical Astrophysics, IAAT, University of T\"ubingen, 72076 T\"ubingen, Germany}
    \affiliation{Department of Physics, University of New Hampshire, 9 Library Way, Durham, NH 03824, USA}
\author{Kostas D. Kokkotas}
    \email{kostas.kokkotas@uni-tuebingen.de}
    \affiliation{Theoretical Astrophysics, IAAT, University of T\"ubingen, 72076 T\"ubingen, Germany}

\date{\today}

\begin{abstract}
We describe a set of time evolution equations and its numerical implementation for the investigation of non-axisymmetric oscillations of rapidly rotating compact objects in full general relativity, taking into account the contribution of a dynamic spacetime. We derive the perturbation equations for the spacetime in the Hilbert gauge, while the hydrodynamical evolution is based on perturbations of the energy-momentum tensor. In our numerical implementation, we use Kreiss-Oliger dissipation in order to achieve a stable time evolution. Our code features high accuracy at comparably low computational expense and we are able to extract the frequencies of non-axisymmetric modes of compact objects with rotation rates up to the Kepler limit.
\end{abstract}

\maketitle

\section{Introduction}

The systematic study of non-axisymmetric neutron star oscillations began in the 1960s with the pioneering work of Thorne and collaborators \cite{1967ApJ...149..591T, 1969ApJ...158....1T} (and subsequent papers), in which they laid out the equations governing the perturbations of a perfect fluid. The numerical solution of these equations has proven highly challenging and it has taken nearly two decades before Lindblom \& Detweiler found an advantageous formulation of the problem as an eigenvalue problem in 1983 that allowed them to numerically integrate the perturbation equations and determine the real and imaginary parts (the frequency and the damping time) of the acoustic modes of sufficiently realistic stellar models \cite{1983ApJS...53...73L, 1985ApJ...292...12D}. These results did not conclude the investigation of these modes; in particular Chandrasekhar \& Ferrari turned to the perturbations of perfect fluids and shed further light on the structure of the equations and their solutions \cite{1991RSPSA.432..247C, 1991RSPSA.434..449C} (and subsequent papers). 

At a similar time in the very early 1990s, after investigating a toy model \cite{1986GReGr..18..913K}, Kokkotas \& Schutz applied techniques borrowed from studies of black hole quasi-normal modes (in their case the WKB approximation which they combined with numerical methods) to the solution in the exterior of the star, and proved that the dynamic spacetime of a neutron star exhibits its very own class of modes and christened them $w$-modes \cite{1992MNRAS.255..119K}. The large imaginary parts of the eigenvalues of these modes pose numerical challenges and this strategy of exploiting black hole science was successfully followed further. In particular, the application of the continued fraction method, the Wronskian method or integrating along anti-Stokes lines led to a considerably more accurate determination of $w$-modes (as well as fluid modes) of neutron stars and the discovery of the $w_{\rm II}$-modes \cite{1993PhRvD..48.3467L, 1995MNRAS.274.1039A}. These numerical solutions finally allowed for the first time to establish universal relations involving frequencies and damping times of oscillation modes, enabling the solution of the inverse problem within the so-called asteroseismology, in which one constrains mass and/or radius of the neutron star, and eventually the nuclear equation of state, via the observation of oscillations \cite{1998MNRAS.299.1059A, 2001MNRAS.320..307K, 2015PhRvD..92l4004D, 2019PhRvL.123e1102T}.

With increasing computational power during the 1990s, the first attempts were made to increase the dimensionality of the hitherto (due to the restriction to spherical symmetry) purely radial problem and include time as a second dimension. The first successful time evolutions of perturbations of relativistic neutron stars were reported by Allen \etal{} in 1998 \cite{1998PhRvD..58l4012A}. In contrast to the integration of ordinary differential equations, the evolution of hyperbolic equations in time is very sensitive to potential irregularities in the coefficients (which in realistically modeled neutron stars are, due to physical phase transitions in the matter, very difficult to avoid) and quickly develops numerical instabilities; Ruoff reformulated those equations by means of the ADM-formalism \cite{ADMformalism} and, crucially, introduced a non-uniform radial grid to overcome such instabilities when using realistic equations of state \cite{2000PhDT.......170R, 2001PhRvD..63f4018R}.

All the previously mentioned studies were concerned with non-rotating stars. The inclusion of rotation proves difficult since the extreme rotation rates that neutron stars may (and do) reach do not allow to neglect the star's oblateness which removes the spherical symmetry from the system; this in turn makes the mathematical formulation much more involved. As a first approximation, rotation was treated perturbatively, too, which allowed to consider even rotating stars as spherically symmetric \cite{1967ApJ...150.1005H}. In this so-called \emph{slow-rotation approximation}, the perturbation equations gain considerably in complexity and have been written down in the Regge-Wheeler gauge \cite{1957PhRv..108.1063R} first by Kojima in 1992 \cite{1992PhRvD..46.4289K}. Even though the problem remains one-dimensional, its solution is not straightforward as (among other technicalities) the outgoing-wave boundary condition at infinity is elusive. Notwithstanding, Andersson successfully applied this formalism and discovered that $r$-modes are prone to the so-called CFS-instability (named after their discoverers Chandrasekhar, Friedman, and Schutz) \cite{1970PhRvL..24..611C, 1975ApJ...199L.157F, 1978ApJ...222..281F} at any rotation rate \cite{1998ApJ...502..708A}. There has been continuing effort using the slow-rotation approximation concerning rotational modes \cite{2001PhRvD..63b4019L, 2003PhRvD..68l4010L}, and also employing different gauges \cite{2002MNRAS.332..676R, 2008MNRAS.384.1711V, 2008PhRvD..77b4029P}, but with the pressing need for frequencies of \emph{rapidly} rotating neutron stars, the interest slowly faded.

Even though the slow-rotation approximation has proven fruitful in the understanding of neutron star physics, it is no longer applicable when considering neutron stars at arbitrary rotation rates, which is essential for nascent neutron stars or post-merger configurations in the immediate aftermath of a binary merger. Without the spherical symmetry of the problem, one has to account for at least two spatial dimensions which complicates the equations further and amplifies the computational expense; furthermore, it remains elusive how to formulate the outgoing-wave boundary condition at infinity for the spacetime perturbations in two spatial dimensions which essentially removes the possibility to formulate a corresponding eigenvalue problem. This issue can be circumvented by adopting the \emph{Cowling approximation} \cite{1941MNRAS.101..367C}, in which the spacetime is considered static, also leading to a considerable simplification of the perturbation equations. Ignoring the impact of a dynamic spacetime (which is most severe for the quadrupolar $f$-mode), Yoshida \& Eriguchi computed quadrupolar $f$-mode frequencies of rapidly rotating neutron stars and studied the associated CFS-instability in the late 90s \cite{1997ApJ...490..779Y, 1999ApJ...515..414Y}. Boutloukos \& Nollert revived this approach and investigated the general properties of the spectrum of neutron stars regarding the acoustic and Coriolis-driven modes \cite{2007PhRvD..75d3007B}. As a further step toward a more general relativistic treatment, Yoshida revisited the problem in the \emph{conformal flatness approximation} \cite{2012PhRvD..86j4055Y}. However, with the mathematical difficulties of extending the eigenvalue formulation to include a dynamic spacetime, the focus shifted to solving this problem as a time evolution.

Despite their complexity, the perturbation equations for rapidly rotating relativistic stars have been written down by Priou already in 1992 \cite{1992MNRAS.254..435P}, even though they were not approached numerically at that time. The most straightforward way of dodging the challenges arising from general relativity is avoiding this theory entirely and treating the task in Newtonian theory. While this is obviously a crude approximation to the strong gravitational field of a neutron star, the numerical solution is less involved and may still result in qualitatively valuable results. Jones \etal{} completed this task formulated as a time evolution in 2002 and their code was later extended \cite{2002MNRAS.334..933J, 2009MNRAS.394..730P}. Returning to a general relativistic treatment, Gaertig \& Kokkotas worked in the Cowling approximation and successfully extracted $f$-mode frequencies of arbitrarily fast rotating neutron stars by adding \emph{artificial viscosity} (also known as Kreiss-Oliger dissipation) \cite{kreiss1973methods} to their evolution equations in order to stabilise their time evolutions \cite{PhDGaertig2008, 2008PhRvD..78f4063G}.

During the first decade of the new millennium, substantial advances were made in the time evolution of the unperturbed, non-linear Einstein equations, mostly driven by the aim to simulate compact binary mergers but obviously also applicable to isolated neutron stars. These systems have hardly any symmetries that can be exploited to reduce the complexity of the problem, requiring to carry out the time evolutions on a three-dimensional grid. The upside of which is that essentially no constraints have to be placed on the rotational profile when simulating the dynamics of a neutron star. Such codes have been seen as a promising new approach to the calculation of mode frequencies of rapidly (and differentially) rotating neutron stars and already at the beginning of the decade, the frequencies of axisymmetric modes in the Cowling approximation \cite{2000MNRAS.313..678F, 2001MNRAS.325.1463F} and those of (quasi-)radial modes in full general relativity \cite{2002PhRvD..65h4024F} had been reported. The non-linear codes kept evolving and were used to generate mode frequencies of $f$-modes in the conformal flatness approximation \cite{2006MNRAS.368.1609D} or those of inertial modes in the Cowling approximation \cite{2008PhRvD..77l4019K}. Not much later, the frequencies of non-axisymmetric modes in full general relativity of rapidly rotating neutron stars obtained from fully non-linear simulations have finally been reported by Zink \etal{} in 2010 \cite{2010PhRvD..81h4055Z}. Even though successful, this approach to computing the frequencies of non-axisymmetric modes, however, has not been followed closely, which is also due to the computational expense associated with such numerical simulations and the accompanying limited accuracy.

In this paper, we will present a perturbative approach to the computation of mode frequencies of rapidly rotating neutron stars accounting, for the first time, for a dynamic spacetime. We study perturbations on an axisymmetric background which allows us to reduce the dimensionality of the problem and keep the computational cost low. For this effort, we extend a code previously developed by us, which we used to investigate oscillations of differentially neutron stars in the Cowling approximation \cite{2010PhRvD..81h4019K}. Except for the inclusion of a few spacetime terms, the hydrodynamical part of the problem remains largely unchanged; for the formulation of the evolution equations of the spacetime perturbations we choose the Hilbert gauge as this gauge will immediately lead to a numerically attractive, fully hyperbolic system. After performing several intrinsic code tests such as convergence tests concerning the grid resolution and the amount of artificial viscosity applied as well as examining the violation of the Hilbert gauge, we compute mode frequencies and compare them to published values from non-linear simulations. As this paper focuses on the technical details of the problem, we report the astrophysically relevant result in an accompanying paper~\cite{2020PhRvL.125l1106K}.

Unless otherwise noted, we employ units in which $c=G=M_\odot=1$ throughout this paper.


\section{Mathematical Formulation}
\label{sec:formulation}

We are going to work with the Einstein equations along with the law for the conservation of energy-momentum,
\begin{equation}
    G_{\mu\nu} = 8\pi T_{\mu\nu}
    \quad\text{and}\quad
    \nabla_\mu T^{\mu\nu} = 0,
    \label{eq:Einstein}
\end{equation}
where $G_{\mu\nu}$ is the Einstein tensor and $T_{\mu\nu}$ is the energy-momentum tensor.

We restrict ourselves to the study of the dynamics of small perturbations around an equilibrium configuration which allows us to linearise equations \eqref{eq:Einstein}. We assume an axisymmetric, stationary background configuration for which the metric written in isotropic coordinates takes the form
\begin{align}
    ds^2
        & = g_{\mu\nu}^{(0)} dx^\mu dx^\nu \\
        & = - e^{2\nu} dt^2 + e^{2\psi} r^2 \sin^2 \theta
            (d\varphi - \omega dt)^2 \\
        & \qquad + e^{2\mu} (dr^2 + r^2 d\theta^2).
            \label{eq:metric} \nonumber
\end{align}
Here, $\nu$, $\psi$, $\mu$, and $\omega$ are the four unknown metric potentials, depending only on $r$ and $\theta$.

We model the neutron star to be a perfect fluid without viscosity for which the corresponding energy-momentum tensor  takes the form
\begin{equation}
    T^{\mu\nu} = (\epsilon + p) u^\mu u^\nu + p g^{\mu\nu},
    \label{eq:Energy-Momentum}
\end{equation}
where $\epsilon$ is the energy density, $p$ is the pressure, and $u^\mu$ the 4-velocity of the fluid. The only two non-vanishing components of the 4-velocity are linked via the star's angular rotation rate, $u^\varphi = \Omega u^t$, and by means of the normalisation of the 4-velocity they are given by
\begin{align}
    u^t = \frac{1}{\sqrt{e^{2\nu}
                    - e^{2\psi} \rrsst \left( \Omega-\omega \right)^2}}.
\end{align}
After specifying an equation of state (henceforth EoS), which may be a polytropic or a tabulated one, linking energy density and pressure to each other, we generate uniformly rotating equilibrium configurations using the \texttt{rns}-code \cite{1995ApJ...444..306S, 1998A&AS..132..431N,rns-v1.1, rns-v1.1}.

\subsection{Introducing Perturbations}

We will now introduce time-dependent perturbations\footnote{Note that we assume a harmonic azimuthal dependence for all perturbations, \ie{} $\tilde{Q} (t,r,\theta,\varphi) = Q(t,r,\theta) e^{im\varphi}$ for any perturbation variable $Q$. This results in the reduction of the dimensionality of the problem and a substantial decrease of computational expense.} for which we will derive evolution equations. First, we decompose the metric as
\begin{align}
    g_{\mu\nu}
        & = g_{\mu\nu}^{(0)} + h_{\mu\nu},
\end{align}
where $g_{\mu\nu}^{(0)}$ is the background metric and $h_{\mu\nu}$ its perturbation; we use the background metric to raise and lower the indices of the latter. As we will work in the Hilbert gauge, it will be advantageous to work instead with the trace-reversed metric perturbation, defined by
\begin{align}
    \phi_{\mu\nu}
        & := h_{\mu\nu} - \frac{1}{2} g_{\mu\nu}^{(0)} h,
\end{align}
where $h := {h^\mu}_\mu$ is the trace of the metric perturbations.

The metric perturbations are not unique but possess gauge freedom which can be utilised in different ways. Often, the gauge freedom is used to eliminate some of the spacetime perturbations, e.g. by using the well-known Regge-Wheeler gauge \cite{1957PhRv..108.1063R}, and hence to reduce the number of perturbation equations. In this study, however, we take a different approach (which we will reason below) and opt for the Hilbert gauge, which is the gravitational equivalent to the well-known Lorenz gauge in electromagnetism, specified by
\begin{equation}
    f_\mu := \nabla^\nu \phi_{\mu\nu} = 0.
    \label{eq:hilbert_gauge}
\end{equation}
In the Hilbert gauge, the perturbed Einstein tensor takes the form
\begin{align}
    - 2 \delta G_{\mu\nu}
        & = \square \phi_{\mu\nu}
          + 2 R^\alpha{}_\mu{}^\beta{}_\nu \phi_{\alpha\beta}
          + R \phi_{\mu\nu} \nonumber \\
        & \quad - \left({R^\alpha}_{\mu} \phi_{\nu\alpha}
          + {R^\alpha}_{\nu} \phi_{\mu\alpha}\right)
          - g_{\mu\nu} R^{\alpha\beta} \phi_{\alpha\beta},
          \label{eq:LinEinstein}
\end{align}
where $R^\alpha{}_\mu{}^\beta{}_\nu$, $R^{\alpha\beta}$, and $R$ are the background Riemann tensor, Ricci tensor and scalar curvature, respectively. The advantage of the Hilbert gauge is that the evolution equations for the metric perturbations will take the form of ten coupled wavelike equations (note that in the above expression, the d'Alembert operator, defined with respect to the background metric, is the only differential operator acting on the metric perturbations) while the mixing of temporal and spatial derivatives is avoided. This is in contrast to other common gauge choices \cite{1998PhRvD..58l4012A, 1971NCimB...3..295B, 2001PhRvD..63f4018R, 2002MNRAS.332..676R} or the ones without any gauge choice \cite{1992MNRAS.254..435P} where the field equations split into subsets of hyperbolic and elliptic equations which have to be solved simultaneously. The fully hyperbolic character of the perturbation equations in the Hilbert gauge makes this gauge particularly convenient for the numerical implementation in a time evolution.

Having opted for the Hilbert gauge, we are dealing with ten functions describing the spacetime perturbations. In choosing our perturbed line element, we are inspired by the one suggested by Priou~\cite{1992MNRAS.254..435P}, which has also been used by Stergioulas \& Friedman in the study of zero-frequency $f$-modes in rapidly rotating stars \cite{1998ApJ...492..301S}; we apply it to the trace-reversed metric perturbations and introduce a few small modifications (mostly factors of $r$ and $\sin\theta$) which turn out crucial in avoiding numerical instabilities. We use calligraphic letters to denote the ten spacetime perturbations $\mcA$, $\mcB$, $\mcH$, $\mcK$, $\mcL$, $\mcM$, $\mcP$, $\mcQ$, $\mcW$, and $\mcY$ and write the perturbed metric such that the individual components take the form
\begin{align}
    \phi_{tt}
        & = -2e^{2\nu}\mcH + 2e^{2\psi} \omega r \sin\theta
            \left( e^{2\nu} \mcY + \omega \rst \mcW \right), \label{eq:metricpert_tt} \\
    \phi_{tr}
        & = \mcL + e^{2\psi} \omega \rst \mcA, \label{eq:metricpert_tr} \\
    \phi_{t\theta}
        & = r \mcM + e^{2\psi} \omega \rrst \mcB, \label{eq:metricpert_tth} \\
    \phi_{t\varphi}
        & = - e^{2\psi} r \sin\theta
            \left( e^{2\nu} \mcY + 2 \omega \rst \mcW \right), \label{eq:metricpert_tph} \\
    \phi_{rr}
        & = 2 e^{2\mu} \mcK, \label{eq:metricpert_rr} \\
    \phi_{r\theta}
        & = e^{2\mu} r \mcQ, \label{eq:metricpert_rth} \\
    \phi_{r\varphi}
        & = - e^{2\psi} \rst \mcA, \label{eq:metricpert_rph} \\
    \phi_{\theta\theta}
        & = 2 e^{2\mu} r^2 \mcP, \label{eq:metricpert_thth} \\
    \phi_{\theta\varphi}
        & = - e^{2\psi} \rrst \mcB, \label{eq:metricpert_thph} \\
    \phi_{\varphi\varphi}
        & = 2 e^{2\psi} \rrsst \mcW. \label{eq:metricpert_phph} 
\end{align}
With this particular choice for the metric components, we can spell out the four constraint equations emerging from the Hilbert gauge, Eq.~\eqref{eq:hilbert_gauge}; we show them in Appendix~\ref{app:hilbert_constraints}.

\subsection{Fluid Perturbations}

Our description for the fluid perturbations is inspired by the very convenient formulation that has been developed and successfully used in \cite{2005JPhCS...8...71K, PhDVavoul2007, 2010PhRvD..81h4019K, 2013PhRvD..88d4052D} for studies within the Cowling approximation. The system of equations describing the hydrodynamics is essentially the same as in the cited literature, except that the equations gain numerous terms accounting for the spacetime perturbations. Nonetheless, we will give a brief overview of their derivation here.

The fundamental idea of the approach is to use the perturbations of the energy-momentum tensor as evolution variables, rather than the typical fluid primitives such as pressure or fluid velocity; this results in a significant shortening of the fluid perturbation equations. In order to avoid confusion regarding the definitions of the perturbation variables in comparison to previous studies, we retain the shortcuts defined in the Cowling approximation and write the perturbed energy-momentum tensor as
\begin{align}
    \delta T^{\mu\nu} & = \delta T^{\mu\nu}_{\rm C} - p h^{\mu\nu},
\end{align}
where $\delta T^{\mu\nu}_{\rm C}$ is the perturbation of the energy-momentum tensor obtained in the Cowling approximation and the minus sign in front of the last term is to account for the fact that $h^{\mu\nu} = - g^{\mu\alpha} g^{\nu\beta} h_{\alpha\beta}$.
We will use the six shortcuts
\begin{align}
    Q_1 & := \delta T^{tt}_{\rm C},        \label{eq:def_Q1} \\
    Q_2 & := \delta T^{t\varphi}_{\rm C} r \sin\theta,     \label{eq:def_Q2} \\
    Q_3 & := \delta T^{tr}_{\rm C},        \label{eq:def_Q3} \\
    Q_4 & := \delta T^{t\theta}_{\rm C} r, \label{eq:def_Q4} \\
    Q_5 & := \delta T^{\varphi\varphi}_{\rm C} r^2 \sin^2\theta,  \label{eq:def_Q5} \\
    Q_6 & := \delta T^{rr}_{\rm C}.        \label{eq:def_Q6}
\end{align}
Note that we have, in fact, slightly modified the definitions of $Q_2$, $Q_4$, and $Q_5$ when compared to previous studies using this formalism. By doing so, we remove the degeneracy of the corresponding base vectors at the rotation axis and considerably improve the numerical stability of the time evolution; technically, these modifications merely shift about a few factors of $r$ and $\sin\theta$ in the perturbation equations but do not substantially alter them, which is why we decide to accept the minor deviation in definition from the previous studies without introducing differently named variables.

The fluid primitives, i.e. the perturbations of energy density, pressure and the fluid velocities, can be reconstructed by inverting the definitions of the variables $Q_1$, $Q_2$, $Q_3$, and $Q_4$. Further, it can be shown that both $Q_5$ and $Q_6$ are not independent variables but are rather a linear combination of the other perturbation variables,
\begin{align}
    Q_5 & = Q_5\left( Q_1, Q_2, Q_6 \right), \\
    Q_6 & = Q_6\left( Q_1, Q_2, \mcH, \mcK, \mcP, \mcW, \mcY \right).
\end{align}
We show the full relations for those two variables in Appendix~\ref{app:pert_emt}.

\subsection{Perturbation Equations}
\label{sec:formulation_pert_eq}

We specified 14 perturbation variables that we need to evolve in time: 10 for the spacetime and 4 for the fluid. The evolution equations follow in a very straightforward manner from the perturbed Einstein equations
\begin{align}
    \delta G_{\mu\nu} & = 8 \pi \delta T_{\mu\nu},
    \label{eq:pert_Einstein}
\end{align}
and the perturbed law for the conservation of energy-momentum
\begin{align}
    \delta \left( \nabla_\mu T^{\mu\nu} \right) & = 0.
    \label{eq:pert_conslaw}
\end{align}
We show the former set of equations, governing the dynamics of the spacetime, in Appendix~\ref{app:spacetime} and the latter set, governing the dynamics of the fluid, in Appendix~\ref{app:fluid}.

One special case of the perturbation equations, which we will consider for testing purposes later in more detail, cf. Sec.~\ref{sec:codetests}, is the non-rotating limit. Setting the star's rotation rate to zero, $\Omega = 0$, immediately implies one of the metric potentials to disappear, $\omega = 0$, and that the polar derivative of the all background quantities vanishes, $\partial_\theta = 0$. Furthermore, in non-rotating stars, the oscillation modes of different azimuthal index $m$ are degenerate and we may set $m = 0$ without loss of generality. This results not only in a dramatic shortening of the evolution equations, but the set of the 14 evolution equations also decomposes into two fully decoupled sets of equations: one set for \emph{polar} perturbations described by the variables  $\mcH$, $\mcK$, $\mcL$, $\mcM$, $\mcP$, $\mcQ$, $\mcW$, $Q_1$, $Q_3$, and $Q_4$, and one set for \emph{axial} perturbations described by the variables $\mcA$, $\mcB$, $\mcY$, and $Q_2$. Beside being an obvious test case for our code, the non-rotating limit has been studied to great detail using time evolution codes as well as eigenvalue codes (see, e.g., \cite{1992MNRAS.255..119K, 1998PhRvD..58l4012A, 1999LRR.....2....2K, 2001PhRvD..63f4018R, 2001A&A...366..565K, 2002PhRvD..66j4002A}) and will not offer any new scientific insight to us.

\subsection{Boundary Conditions}

The perturbation equations are complemented by boundary conditions which describe the behaviour of the perturbations on the boundaries of the numerical domain. As we will explain in Sec.~\ref{sec:numerical_methods}, our numerical domain is bounded by the star's rotation axis and the equatorial plane in the polar direction; in the radial direction, we have to apply boundary conditions at the star's surface for the fluid perturbations as well as at the outer edge of the numerical grid for the spacetime perturbations. As our description of the fluid is largely the same as in \cite{2010PhRvD..81h4019K}, most of the boundary conditions for the fluid can be found there but due to slight modifications in the definitions and for the sake of completeness, we will repeat them here.

\begin{itemize}[nosep,leftmargin=0pt,labelindent=\labelsep,itemindent=*]

\item Having the equatorial plane, characterised by $\theta=\pi/2$, as a boundary of our numerical domain allows us to select one of two disjoint sets of oscillations modes that behave differently under reflection about the equatorial plane. Let $R$ be the corresponding reflection operator, which in spherical coordinates is given by the mapping $\left( r,\theta,\varphi \right) \mapsto \left( r,\pi-\theta,\varphi \right)$; it is easy to show that for any spherical harmonic, $Y_{lm}$, the relation $RY_{lm} = (-)^{l-m} Y_{lm}$ holds. In the non-rotating limit, all perturbation variables can be expanded into spherical, vector and tensor harmonics \cite{1957PhRv..108.1063R} and, according to their behaviour under the operator $R$, they can be categorised into the two sets $\bm{R_1} := \left\{ \mcA, \mcH, \mcK, \mcL, \mcP, \mcW, \mcY, Q_1, Q_2, Q_3 \right\}$ and $\bm{R_2} := \left\{ \mcB, \mcM, \mcQ, Q_4 \right\}$. Depending on whether $l-m$ is even or odd,\footnote{We note that we use the words \emph{even} and \emph{odd} only when referring to the integer difference $l-m$; in some studies, polar modes are also called ``even modes'' (and axial ones are called ``odd'') but we do not adopt this nomenclature to avoid confusion.} the perturbation variables of one of those two sets are either zero at the equatorial plane or their $\theta$-derivative vanishes there; we show a summary in Table~\ref{tab:bc_eqplane}.

\item To derive the boundary conditions at the rotation axis, which is given by $\theta = 0$, we will again utilise the angular behaviour of the perturbation variables as specified by the spherical harmonics and their tensorial equivalents. In this instance, we have to evaluate the limits of the perturbations as $\theta$ approaches 0 for each azimuthal index $m$; however, we will restrict our study to the astrophysically relevant cases of $m \ge 2$. We find that for $m \ge 3$, all perturbations vanish along the rotation axis, while for $m = 2$ they separate into the set of perturbations $\bm{\Theta_1} := \left\{ \mcB, \mcP, \mcW \right\}$, whose $\theta$-derivative vanishes at the rotation axis, and the set of perturbations $\bm{\Theta_2} := \left\{ \mcA, \mcH, \mcK, \mcL, \mcM, \mcQ, \mcY, Q_1, Q_2, Q_3, Q_4 \right\}$, which become zero along the rotation axis. We show a summary of the boundary conditions at the rotation axis in Table~\ref{tab:bc_rotaxis}.

\item The origin is mapped onto a boundary line of our numerical grid. Physically, however, it remains part of the rotation axis and the equatorial plane. When a perturbation variable is shown to vanish on either of the neighbouring boundary lines, we set it to zero at the origin as well; otherwise, we use the same value as on the next grid point.

\item At the stellar surface, we need to apply boundary conditions to the fluid perturbations. The variables $Q_3$ and $Q_4$ vanish by virtue of their definition
\begin{align}
    Q_3 = \delta T^{tr}_{\rm C} = \left( \epsilon + p \right) u^t \delta u^r = 0, \\
    Q_4 = \delta T^{t\theta}_{\rm C} r = \left( \epsilon + p \right) u^t \delta u^\theta r = 0,
\end{align}
because pressure and energy density vanish there. $Q_1$ and $Q_2$ merely need to exhibit ``numerically smooth'' behaviour there and we extrapolate them linearly in radial direction from the interior of the star.

For the case of rotating and hence oblate stars, the stellar surface traverses our grid. As an approximation to the true stellar surface, we apply the boundary conditions for the fluid variables at the very first grid point (counting from the origin) that is located outside the star; see also Fig.~\ref{fig:grid_star}.

\item Last, we consider the outer edge of the numerical grid which is typically located about a hundred or so star radii away from the origin and can be chosen arbitrarily; there is no physical boundary or mathematical singularity at this location. We employ a Sommerfeld boundary condition at this edge in order to remove the energy carried by gravitational waves from the grid. While it accounts correctly only for spherical wave fronts but not for waves traveling in arbitrary directions, this condition offers a highly attractive ratio of simplicity of implementation when compared to the amount of artificially reflected energy and has also been used in 3-dimensional simulations \cite{1999PhRvD..59b4007B, 2000PhRvD..62d4034A}; furthermore, at the distances mentioned above, a spherical wave front will be a very good approximation to the outgoing radiation for our relevant scenarios. This is confirmed in our simulations, in which we observe only marginal reflection that does not spoil the later stages of our time evolutions.

\end{itemize}

\begin{table}[htpb]
    \centering
    \caption{Boundary condition of the perturbation variables at the equatorial plane, $\theta = \pi/2$. The two sets $\bm{R_1}$ and $\bm{R_2}$ are defined in the text and the stated boundary condition applies to each member of those sets.}
    \label{tab:bc_eqplane}
    \begin{tabular}{c|c|c}
        \toprule
         $l-m$ & $\bm{R_1}$ &  $\bm{R_2}$ \\
        \midrule
        even & $\partial_\theta = 0$ & $0$ \\
        odd  & $0$ & $\partial_\theta = 0$ \\
        \bottomrule
    \end{tabular}
\end{table}

\begin{table}[htpb]
    \centering
    \caption{Boundary conditions of the perturbation variables along the rotation axis, $\theta = 0$. The two sets $\bm{\Theta_1}$ and $\bm{\Theta_2}$ are defined in the text and the stated boundary condition applies to each member of those sets.}
    \label{tab:bc_rotaxis}
    \begin{tabular}{c|c|c}
        \toprule
         & $\bm{\Theta_1}$ & $\bm{\Theta_2}$ \\
        \midrule
        $m = 2$   & $\partial_\theta = 0$ & 0 \\
        $m \ge 3$ & $0$                   & \makebox*{$\partial_\theta = 0$}{$0$} \\
        \bottomrule
    \end{tabular}
\end{table}

We need to comment further on the boundary conditions in polar direction, \ie those applied at the equatorial plane and at the rotation axis; we have derived those from the angular behaviour of the spherical harmonics which requires the background configuration to be spherically symmetric. While it has been shown that those boundary conditions are valid also in the slow-rotation approximation~\cite{1992PhRvD..46.4289K, 2005IJMPD..14..543S}, it is not immediately clear that they also hold in rapidly rotating stars.

The equatorial plane remains a plane of symmetry also when the star is oblate. As none of the two hemispheres is distinguished, the oscillation modes remain eigenfunctions of our previously defined reflection operator $R$. Further, it has been argued within the slow-rotation formalism that all oscillation modes, even though gaining contributions from other modes, keep their overall parity~\cite{1999ApJ...521..764L, 2001PhRvD..63b4019L}; hence, their eigenvalue with respect to $R$ remains unchanged. Further, despite the simplification of the slow-rotation approximation, the relevant couplings between the different mode classes is already present and it remains the same in rapidly rotating stars; thus, the argument can be extended to the case of rapid rotation without further ado, leaving the boundary conditions unaltered.

In order to understand the boundary conditions at the rotation axis without relying on the star's spherical symmetry, we study the behaviour of the perturbation variables under rotation about the rotation axis, $(r, \theta, \varphi) \mapsto (r, \theta, \varphi')$. All of our evolution variables describe scalar, vectorial or tensorial (of rank 2) perturbations; in order for them to be well-defined at the rotation axis, they may have a non-zero value only if their azimuthal dependence resembles that which is specified by the azimuthal parameter $m$. Hence, for $m \ge 3$ all perturbations have to vanish along the rotation axis; when $m = 2$, all scalar and vectorial perturbations have to vanish there, too, and only the tensorial components (that have rank 2) of the perturbed metric, i.e. $\mcB$, $\mcP$, and $\mcW$, may be non-zero. Instead, it is their polar derivative which has to vanish there, i.e. $\partial_\theta = 0$, to ensure smoothness of the solution across the poles.


\section{Numerical Methods}
\label{sec:numerical_methods}

After decomposing the perturbed quantities with respect to $\varphi$, the resulting evolution equations form a two-dimensional problem in the spherical coordinates $r$ and $\theta$. We evolve the perturbations on a grid with radial coordinate $s \in \left[ 0, 1 \right]$ and angular coordinate $t \in \left[ 0, 1 \right]$; the mapping to the spherical coordinates $r$ and $\theta$ is given by
\begin{align}
    r & = \frac{b r_e \sqrt{s}}{a - \sqrt{s}}, \\
    \theta & = \arccos t,
\end{align}
where $r_e$ is the equatorial coordinate radius of the star and $a$ and $b$ are two parameters with which we can adjust the radial grid.

\begin{figure}[htbp]
    \centering
    \includegraphics[width=8.6cm]{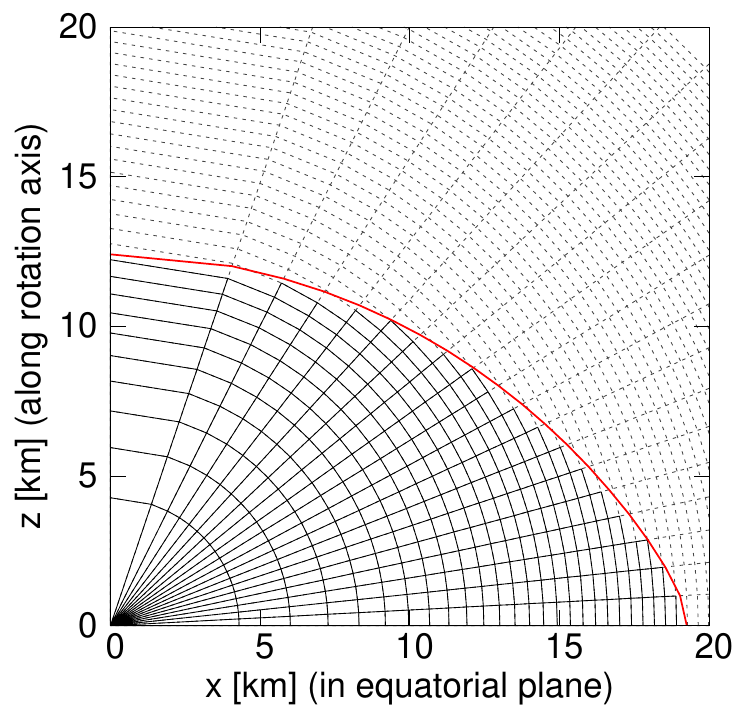}
    \caption{The computational grid used for the fast rotating model BU8, which we will discuss in Sec.~\ref{sec:codetests_rotating}. The red line depicts the surface of the star as obtained from the \texttt{rns}-code; the solid black lines depict the computational grid inside the star, while the grey dashed lines depict the grid in the exterior of the star. The boundary conditions for the stellar surface are applied at the grid points just ``outside'' the red line.}
    \label{fig:grid_star}
\end{figure}

This particular radial grid function, $r(s)$, has a varying grid spacing across the entire radial domain and allows us to avoid the introduction of different grids for different parts of the radial domain and grid interpolation between those different grids. In particular, the grid spacing is comparably coarse near the centre of the star, becomes much finer in the region around the star's surface, and then becomes coarser again in the exterior of the star where it approaches a near uniform grid spacing at the outer edge; we show a sketch of our computational grid in the vicinity of a fast rotating star in Fig.~\ref{fig:grid_star}. Furthermore, it comes with two parameters $a$ and $b$ with which we can steer the behaviour of the grid; in particular, we choose $a$ and $b$ such that the outer edge of the grid is roughly a hundred star radii away from the star and that the physical grid spacing in radial direction is limited by half the star's radius in that region. We find that this grid offers a reasonable compromise between having a grid extending sufficiently into the far field, while keeping the number of grid points and hence the computational expense low.

The grid coordinates are not fitted to the shape of the star, which means that the surface of rotating and thus oblate stars lies in between the grid points of our computational domain. We approximate the surface of the star to be located at those grid points for which the energy density first vanishes when moving outward from the centre along lines of constant polar angle; this is where we apply the surface boundary condition for the fluid variables.

We use central differences of second order to discretise the spatial derivatives. For the time evolution, we use the method of lines and apply the third-order Runge-Kutta method as time integrator. Similar to previous studies within the same perturbative framework \cite{2008PhRvD..78f4063G, 2010PhRvD..81h4019K}, this code is prone to numerical instabilities. As the hydrodynamical equations are essentially identical to those in \cite{2010PhRvD..81h4019K}, we also encounter the same instability near the origin of the star; we apply second-order Kreiss-Oliger dissipation \cite{kreiss1973methods} to cure those instabilities. The spacetime perturbations do not suffer from similar instabilities near the origin as potential instabilities emerging from truncation errors will be radiated away; however, we observe some unstable behaviour in the spacetime perturbations in the far field where the grid becomes coarser. These instabilities stem from the fact that our grid has difficulties resolving the gravitational waves in the wave zone. We could increase the grid resolution to overcome this issue; however, we find that applying a very small amount (typically 10--100 times smaller than that for the fluid) of Kreiss-Oliger dissipation to the spacetime grid eliminates those instabilities, too. Furthermore, even an increased grid resolution cannot entirely prevent this issue since gravitational waves of sufficiently high frequency will confront us with the same difficulties.

The typical grid resolution that we employ for our simulations is $3120 \times 50$ points, where the radial grid extends to $100 R_e$. The interior of the star is then covered by $181 \times 50$ grid points. For this resolution, we vary the dissipation coefficients for the fluid between $\approx 1 \times 10^{-7} - 5 \times 10^{-6}$, even though we distinguish between dissipation coefficients for the radial and for the angular direction (in some cases a larger dissipation in angular direction is required); from our experience, the amount of dissipation required to achieve a stable numerical simulation depends on the background model and varies with central energy density and rotation rate. The spacetime perturbations behave in a more benign manner and we leave the dissipation coefficient of $10^{-8}$ largely untouched. In general, finding those dissipation coefficients that are ``optimal'' in the sense that the least amount of artificial viscosity is closest to the physical reality (ignoring the fact that we are dealing with highly idealised neutron star models) is within certain bounds a matter of trial and error.

In order to extract the mode frequencies, we observe the temporal evolution of one of our perturbations variables at one freely chosen point on our grid and then calculate the discrete Fourier transform of this time series. A considerable improvement of the obtained frequencies can be achieved by locating the roots of the derivative of the power spectral density, which we approximate with second order central finite differences.


\section{Code Tests}
\label{sec:codetests}

We have described the mathematical formulation of the problem as well as the numerical methods that we have employed to solve the problem. It is crucial for a newly developed code to be thoroughly tested. We will start by showing results from several convergence tests for various modes of non-rotating and rotating configurations; along with those tests, we will list for future reference the $f$-mode frequencies of the well-known BU sequence which has served as a popular test case in various studies \cite{2004MNRAS.352.1089S, 2006MNRAS.368.1609D, 2008PhRvD..77l4019K, 2008PhRvD..78f4063G}. Next, we will turn to our treatment of the Hilbert gauge and how its violation evolves throughout our simulations. Last, we will reproduce some mode frequencies and compare them to previously published data.

\subsection{Convergence Test}

\subsubsection{Non-Rotating Neutron Star Model}

Our very first convergence test is for the $f$-mode of the regularly employed BU0 model, which is a simple polytropic model with no spin, characterised by the polytropic constants $N=1$, $K=100$ and a central energy density of $\epsilon_c = 0.891 \cdot 10^{15} \unit{g/cm}^3$; these parameters result in a neutron star model with typical values of $M = 1.400 M_\odot$ and $R_e = 14.16\unit{km}$. Even though this model has previously been studied extensively, we performed a detailed convergence test in order to understand the general behaviour of our code, not impacted by potential numerical challenges stemming from the oblateness of rapidly rotating models or tabulated EoSs.

We perform a series of simulations using several different grid resolutions; starting from our ``standard'' grid resolution of $3120 \times 50$ points, we decrease or increase the number of grid points in uniform steps. Since we have chosen a highly non-uniform radial grid (cf. Sec.~\ref{sec:numerical_methods}), we show some of its characteristics in Table~\ref{tab:grid_charac}; it is apparent that the grid spacing or the number of points inside the star is not proportional to the number of grid points in radial direction, but this does not pose a problem as long as the map between them is strictly monotonic. We emphasise that we show physical distances relating to the evolution grid of the BU0 model in Table~\ref{tab:grid_charac} for intuition only; however, they are not universal and depend on the particular neutron star model, in particular its oblateness; they scale with the star's equatorial radius, $R_e$, and hence vary only moderately.

\begin{table}[htpb]
    \centering
    \caption{We show some characteristic values of our radial grid for different grid resolutions. It is in the nature of our grid function that the part of the grid that covers the star depends on the overall number of grid points ($n_{\rm star}$ is the number of grid points inside the star along the equatorial plane). In particular, the radial grid spacing at the origin, $\Delta r(0)$, is comparably large, while the radial spacing at the surface, $\Delta r(R_e)$, is very fine, which allows a good resolution of the star's surface. For all listed grids, the outer edge of the grid is located at $100R_e$ and the radial spacing at this edge is $5950\unit{m}$. The values shown here are indicative and depend mildly on the oblateness of the star.}
    \label{tab:grid_charac}
    \begin{tabular}{crrr}
        \toprule
        Grid points & $\Delta r(0)$ & $\Delta r(R)$ & $n_{\rm star}$ \\
        \midrule
        $3900 \times 62$ & $713\unit{m}$ & $26\unit{m}$ & 312 \\
        $3120 \times 50$ & $998\unit{m}$ & $43\unit{m}$ & 180 \\
        $2496 \times 40$ & $1395\unit{m}$ & $73\unit{m}$ & 102 \\
        $1997 \times 32$ & $1949\unit{m}$ & $126\unit{m}$ & 57 \\
        \bottomrule
    \end{tabular}
\end{table}

\begin{figure}[htbp]
    \centering
    \includegraphics[width=8.6cm]{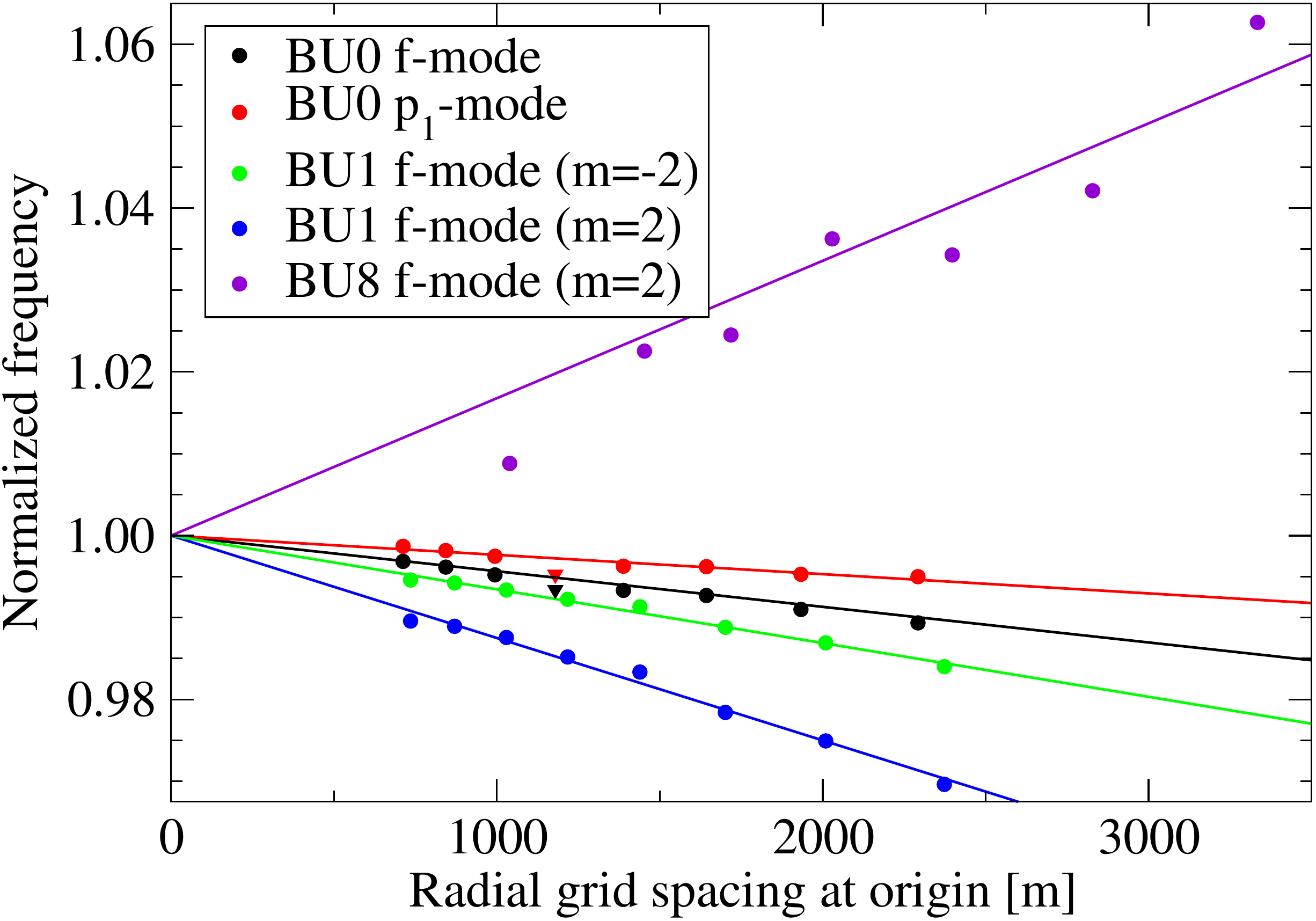}
    \caption{Graphical visualisation of the convergence tests for the five selected oscillation modes discussed in the text. The frequencies of the modes are normalised by their respective limiting value; the straight lines are linear fits to the data. The results clearly show linear convergence of the code and even for fairly coarse grids, the frequencies are recovered with only a few percent error. The two data points depicted by triangles are removed from the linear fits as outliers. For actual data refer to Tables~\ref{tab:convtest_BU0_2f}, \ref{tab:convtest_BU0_2p1}, and \ref{tab:convtest_BU1_BU8}.} 
    \label{fig:conv_test}
\end{figure}

When comparing obtained frequencies for different grid resolutions, we have to take into account the necessity for artificial viscosity which will impact the physical observables to a small extent. For the BU0 model, we find that the lower bound for the dissipation coefficients is somewhere around $10^{-7}$; we run the simulations for different grid resolutions as well as for three different fluid dissipation coefficients and show the resulting frequencies of the $f$-mode and the $p_1$-mode for $l=2$ in Tables~\ref{tab:convtest_BU0_2f} and \ref{tab:convtest_BU0_2p1}; further, we show the frequencies of both modes and linear fits in Fig.~\ref{fig:conv_test}. We want to make a note on the accuracy of the listed frequencies: In addition to locating the roots of the derivative of the power spectral density, we also analyse smaller chunks of the time series. By varying size and position of the time series window, we obtain marginally different values (due to spectral leakage) over which we then average; this procedure results in a precision of well below $1\unit{Hz}$. For this method to work, it is crucial that we ensure the surface of the star to match up with an angular grid line (the location at which we apply the surface boundary conditions); this requires a meticulous fine-tuning of the grid parameters and is in this way, unfortunately, not possible for rotating stars as their surface traverses the grid in between the grid points. Without this fine-tuning of the grid, the obtained frequencies would appear to randomly deviate by several Hz from the listed ones and not result in a meaningful convergence graph.

\begin{table}[htpb]
    \centering
    \caption{Frequencies of the ${}^2f$-mode of the non-rotating BU0 model for different grid resolutions (the grid resolution labeled ``$\infty$'' denotes extrapolation to the continuum limit) and different dissipation coefficients for the fluid. The dissipation coefficient for the spacetime is fixed at $10^{-8}$. The simulation duration is $15\unit{ms}$. All frequencies are given in kHz.}
    \label{tab:convtest_BU0_2f}
    \begin{tabular}{cccc}
        \toprule
        Grid points &  \multicolumn{3}{c}{Dissipation coefficient} \\
        \cmidrule(lr){2-4}
         & $5 \cdot 10^{-7}$ & $7 \cdot 10^{-7}$ & $10 \cdot 10^{-7}$ \\
        \midrule
        $\infty$         & $1.5859$ & $1.5858$ & $1.5855$ \\
        $3900 \times 62$ & $1.5807$ & $1.5808$ & $1.5809$ \\
        $3488 \times 56$ & $1.5797$ & $1.5797$ & $1.5798$ \\
        $3120 \times 50$ & $1.5780$ & $1.5782$ & $1.5784$ \\
        $2790 \times 45$ & $1.5747$ & $1.5749$ & $1.5750$ \\
        $2496 \times 40$ & $1.5749$ & $1.5752$ & $1.5755$ \\
        $2232 \times 36$ & $1.5739$ & $1.5742$ & $1.5746$ \\
        $1997 \times 32$ & $1.5711$ & $1.5715$ & $1.5721$ \\
        $1786 \times 29$ & $1.5683$ & $1.5689$ & $1.5698$ \\
        \bottomrule
    \end{tabular}
\end{table}

\begin{table}[htpb]
    \centering
    \caption{Same as Table~\ref{tab:convtest_BU0_2p1} but for the ${}^2p_1$-mode.}
    \label{tab:convtest_BU0_2p1}
    \begin{tabular}{cccc}
        \toprule
        Grid points &  \multicolumn{3}{c}{Dissipation coefficient} \\
        \cmidrule(lr){2-4}
         & $5 \cdot 10^{-7}$ & $7 \cdot 10^{-7}$ & $10 \cdot 10^{-7}$ \\
        \midrule
        $\infty$         & $3.7220$ & $3.7227$ & $3.7236$ \\
        $3900 \times 62$ & $3.7185$ & $3.7186$ & $3.7187$ \\
        $3488 \times 56$ & $3.7169$ & $3.7169$ & $3.7168$ \\
        $3120 \times 50$ & $3.7140$ & $3.7142$ & $3.7142$ \\
        $2790 \times 45$ & $3.7061$ & $3.7062$ & $3.7059$ \\
        $2496 \times 40$ & $3.7107$ & $3.7102$ & $3.7097$ \\
        $2232 \times 36$ & $3.7110$ & $3.7105$ & $3.7095$ \\
        $1997 \times 32$ & $3.7075$ & $3.7071$ & $3.7060$ \\
        $1786 \times 29$ & $3.7076$ & $3.7063$ & $3.7049$ \\
        \bottomrule
    \end{tabular}
\end{table}

When performing the convergence test for both the $f$-mode and the $p_1$-mode, we observe that the frequencies obtained using the $2790 \times 45$ grid deviate somewhat from an otherwise pretty linear convergence behaviour displayed for all other grid resolutions. At this time, it is not clear to us why this happens and we exclude the results from those simulations from the convergence test (but nonetheless show them in the graph, cf. Fig.~\ref{fig:conv_test}). Further, it is obvious that the impact of the artificial dissipation on the extracted oscillation frequencies is minuscule and can safely be ignored. Overall, the convergence test demonstrates convergence of our code and by extrapolation, we yield the frequencies $1.586\unit{kHz}$ for the $f$-mode and $3.723\unit{kHz}$ for the $p_1$-mode. Those frequencies deviate only by $0.5\%$ from the frequencies obtained by a (naturally much more precise) eigenvalue code: they are $1.5780\unit{kHz}$ and $3.7068\unit{kHz}$, respectively. As both convergence tests are consistent by themselves, the remaining deviation must be a systematic error, stemming from numerical inaccuracies: the most prominent candidate for error is the fact that we are using two two-dimensional codes (the \texttt{rns}-code for the background configuration and our time evolution code) to tackle a problem which, due to its spherical symmetry, is inherently one-dimensional; while this is mathematically not a problem, we are limited by the computational expense to somewhat lower accuracy. Furthermore, we will inevitably introduce some small interpolation error when transferring the background quantities from the grid of the \texttt{rns}-code to our computational grid.

\subsubsection{Rotating Neutron Star Model}
\label{sec:codetests_rotating}

We continue with convergence tests for rotating configurations. Having begun with the non-rotating model BU0 of the BU sequence in the previous section, we now turn to its rotating members. Along this sequence, the central energy density is kept fixed at $\epsilon_c = 0.891 \cdot 10^{15} \unit{g/cm}^3$ while the axis ratio $r_p/r_e$ is lowered in uniform steps of 0.05 until the Kepler limit is reached \cite{2004MNRAS.352.1089S}; this happens at an axis ratio of $r_p/r_e = 0.576$ and we call this model BUK.

We perform convergence tests for the $f$-mode frequency of the model BU1, which is the slowest rotating member of the BU sequence, and the model BU8, which is very close to the mass-shedding limit. As explained before, it is not possible for models of rotating stars to have their surface perfectly aligned with the grid lines (cf. also Fig.~\ref{fig:grid_star} for the computational grid employed for the BU8 model); nonetheless, the impact of the non-alignment of the surface with the grid on the mode frequencies seems to even out over the angular direction and we observe the frequencies converging with increasing grid resolution. We list the obtained frequencies in Table~\ref{tab:convtest_BU1_BU8} and show them also in Fig.~\ref{fig:conv_test}. We keep the dissipation coefficient for the spacetime fixed at $10^{-8}$; for the fluid, the dissipation coefficients is $10^{-6}$. As in the convergence test for the BU0 model, we observe only marginal dependence of the frequencies on the artificial dissipation and hence do not show those results separately.

\begin{table}[htpb]
    \centering
    \caption{The convergence test for both $l=|m|=2$ $f$-modes of the BU1 model (with a simulation time of $15\unit{ms}$) and for the $l=m=2$ $f$-mode of the BU8 model (simulation time of $30\unit{ms}$). The dissipation coefficient for the fluid was $10^{-6}$. All frequencies are given in kHz.}
    \label{tab:convtest_BU1_BU8}
    \begin{tabular}{cccc}
        \toprule
        Grid points & \multicolumn{2}{c}{BU1} & BU8 \\
        \cmidrule(lr){2-3} \cmidrule(lr){4-4}
        & $\sigma({}^2f_2)$ & $\sigma({}^2f_{-2})$ & $\sigma({}^2f_2)$ \\
        \midrule
        $\infty$         & $1.0926$ & $1.9547$ & $-0.1021$ \\
        $3900 \times 62$ & $1.0812$ & $1.9441$ & $-0.1030$ \\
        $3488 \times 56$ & $1.0805$ & $1.9434$ & $-0.1059$ \\
        $3120 \times 50$ & $1.0790$ & $1.9417$ & $-0.1044$ \\
        $2790 \times 45$ & $1.0764$ & $1.9395$ & $-0.1046$ \\
        $2496 \times 40$ & $1.0744$ & $1.9377$ & $-0.1058$ \\
        $2232 \times 36$ & $1.0690$ & $1.9328$ & $-0.1056$ \\
        $1997 \times 32$ & $1.0652$ & $1.9291$ & $-0.1064$ \\
        $1786 \times 29$ & $1.0594$ & $1.9234$ & $-0.1085$ \\
        \bottomrule
    \end{tabular}
\end{table}

Our convergence tests prove that our code allows us to determine the frequencies of oscillation modes to an accuracy of around $1\%-2\%$ or better when sticking to the resolution of $3120 \times 50$ grid points. We list the frequencies of the $l=|m|=2$ $f$-mode for all members of the BU sequence in Table~\ref{tab:freq_BU_sequence}. Compared to the values in the Cowling approximation (cf. Refs. \cite{2008PhRvD..78f4063G,2010PhRvD..81h4019K}), the frequencies in full general relativity are, as expected, lower: for this particular sequence, the frequencies of both branches experience a shift of roughly $200-300\unit{Hz}$ with the exception of the fastest rotating models BU8 and beyond; for those, the shift in frequency for the counter-rotating mode is closer to $\approx 100\unit{Hz}$. This results in a slightly enhanced parameter window for the CFS-instability.

\begin{table}[htpb]
    \centering
    \caption{Frequencies of the counter- and co-rotating $f$-mode of the BU sequence as obtained from simulations on a grid with $3120 \times 50$ points (we have extended the sequence by the two models BU8a and BUK). We show the axis ratio, $r_p/r_e$, for completeness. The simulation time is usually between $15$ and $30\unit{ms}$; to achieve the displayed accuracy of the ${}^2f_2$-mode of the BU7 model, we evolved this particular model for $300\unit{ms}$ (and obtained $11.3\unit{Hz}$). All frequencies are given in kHz.}
    \label{tab:freq_BU_sequence}
    \begin{tabular}{ld{3.3}d{3.3}c}
        \toprule
        Model & \mc{$r_p/r_e$} & \mc{$\sigma({}^2f_2)$} & $\sigma({}^2f_{-2})$ \\
        \midrule
        BU0  & 1.00 & 1.578 & $1.578$ \\
        BU1  & 0.95 & 1.079 & $1.942$ \\
        BU2  & 0.90 & 0.836 & $2.054$ \\
        BU3  & 0.85 & 0.636 & $2.118$ \\
        BU4  & 0.80 & 0.456 & $2.155$ \\
        BU5  & 0.75 & 0.293 & $2.172$ \\
        BU6  & 0.70 & 0.146 & $2.182$ \\
        BU7  & 0.65 & 0.011 & $2.188$ \\
        BU8  & 0.60 & -0.104 & $2.199$ \\
        BU8a & 0.59 & -0.124 & $2.200$ \\
        BU9  & 0.58 & -0.138 & $2.201$ \\
        BUK  & 0.576 & -0.140 & $2.201$ \\
        \bottomrule
    \end{tabular}
\end{table}

\subsection{Hilbert Gauge Violation}

In the derivation of the evolution equations for the spacetime perturbations, we have made use of the Hilbert conditions, cf.~Eq.~\eqref{eq:hilbert_gauge}, in order to simplify the lengthy equations. Nonetheless, the four gauge conditions, $f_\mu = 0$, supplement the ten field equations and should be fulfilled at all times, even though the latter are sufficient to evolve the system in time (we are ignoring the fluid perturbations for the moment as they are not affected by the gauge choice). The apparent overdetermination of the system is remedied by the fact that the field equations preserve the satisfaction of the gauge conditions if they are satisfied on the initial time slice. That this holds true has been shown for vacuum spacetimes by Barack \& Lousto \cite{2005PhRvD..72j4026B}. In the presence of neutron star matter, the proof is a little more involved but yields a comparable result. Since the divergence of the Einstein tensor vanishes, the perturbation thereof must vanish, too, $\delta\left( \nabla^\nu G_{\mu\nu} \right) = 0$. By virtue of the contracted Bianchi identities, it then follows that this condition is equivalent to the homogeneous wave equation
\begin{equation}
    \square f_\mu + {R^\nu}_\mu f_\nu = 0
\end{equation}
for the divergence of $\phi_{\mu\nu}$. Hence, if we pick initial data such that the Hilbert conditions are fulfilled on the initial time slice and that their first time derivative vanishes, $\left( \partial_t f_\mu \right)\rvert_{t=0} = 0$, they will remain satisfied throughout the evolution.

\begin{figure}[htbp]
    \centering
    \includegraphics[width=8.6cm]{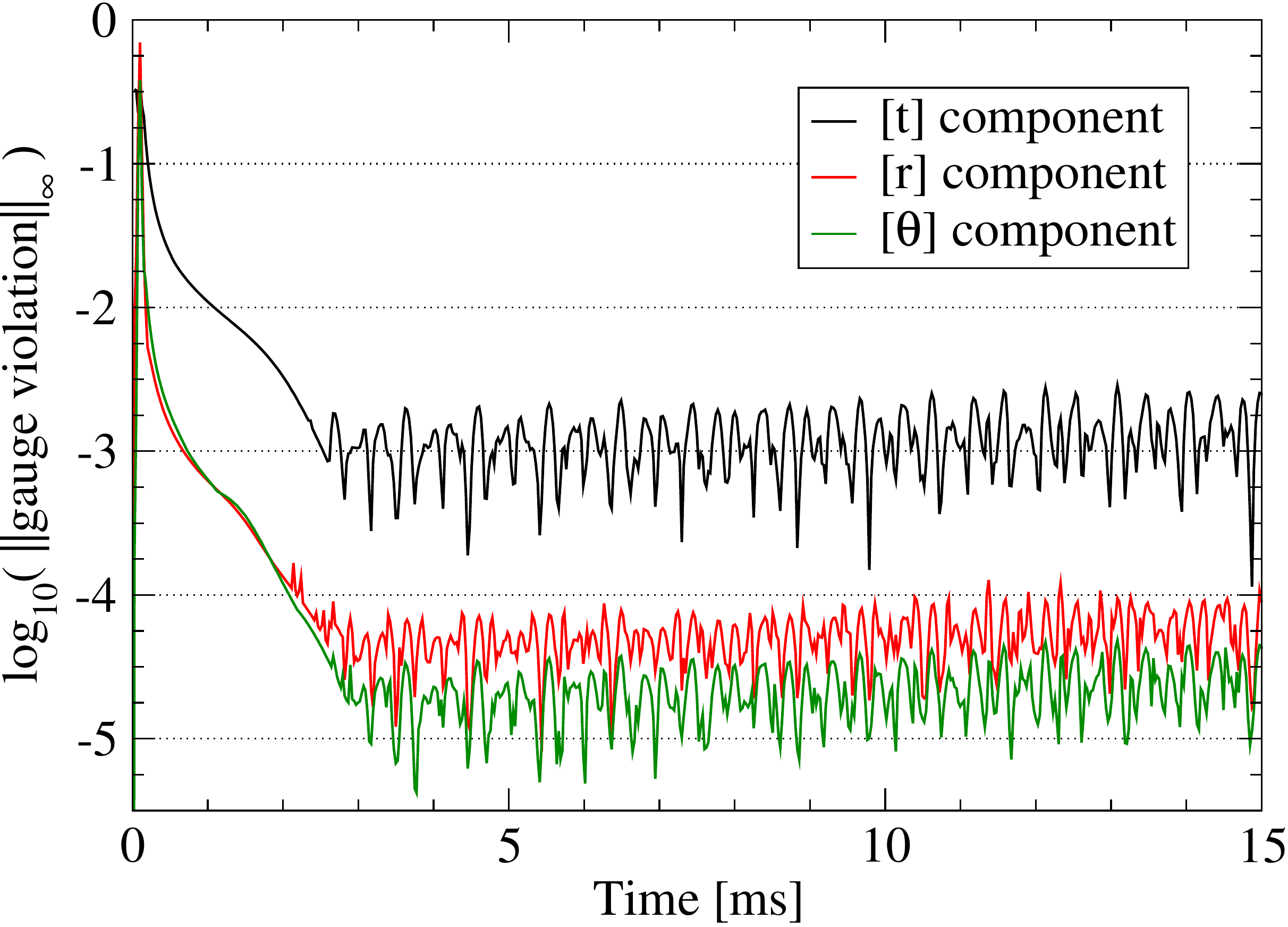}
    \caption{The numerical violation of the Hilbert gauge. We show the maximum norm of the three non-zero Hilbert conditions, $f_\mu = 0$, during the time evolution of a polar perturbation of a non-rotating star. The gauge violation clearly remains bounded. The apparently severely amplified gauge violation during the first two milliseconds can be justified by noting that we show the absolute violation rather than the one relative to the perturbation amplitude (see text for further explanation).}
     \label{fig:hilbert-violation}
\end{figure}

Clearly, the previous statement holds true only in the continuum limit and the gauge conditions will inevitably be violated in our simulations due to various truncation and round-off errors in the numerical implementation. There is the possibility that such gauge violations grow out of control over the course of the time evolution and render it void of physical meaning. However, we do not observe such destructive behaviour in our simulations; in contrast, the gauge violation remains bounded without further ado. Furthermore, we observe that the violation decreases with increasing grid resolution. We show the evolution of the Hilbert gauge violations in Fig.~\ref{fig:hilbert-violation} during the simulation of a polar perturbation of a non-rotating star. More precisely, the graph shows the maximum norm of the three non-zero gauge conditions across the time slices; the three gauge violations clearly remain bounded by $3\cdot 10^{-3}$ and $10^{-4}$, respectively.\footnote{As explained in Sec.~\ref{sec:formulation_pert_eq}, the problem decomposes into two disjoint sets of equations in the case of no rotation and for polar perturbations only three of the four Hilbert conditions are non-zero.} The apparently severely amplified gauge violation during the first half of a millisecond of the evolution (after it started at machine precision in the initial data) can be justified by noting that we show the absolute violation rather than the one relative to the perturbation amplitude; as our initial Gaussian pulse travels toward the origin of our spherical coordinate system at the beginning of the simulation, the Gaussian pulse grows in amplitude and so does the absolute violation of the Hilbert conditions. As the graph clearly shows, this comparatively large violation is only temporary and once the energy is radiated away, it reduces to numerically expected levels.

We now turn to the Hilbert conditions concerning rotating configurations. The mathematical complexity of the problem escalates as rotation sets in, and then the Hilbert conditions (and their first time derivatives) form an involved and strongly coupled set of partial differential equations, for which we were not able to construct a straightforward appropriate solution. However, as we are currently concerned with the dynamics of fluid perturbations which are not affected by the gauge choice, a potential violation of the Hilbert conditions is irrelevant for the observables of interest. This is confirmed by the fact that, first, we observe the identical spectrum for non-rotating stars, irrespective of whether or not we choose to satisfy the gauge conditions on the initial time slice, and second, we find excellent agreement between our and previously published results (cf. Sec.~\ref{sec:codetests_comparison}).

A numerical solution satisfying the Hilbert gauge (to numerically acceptable precision) is clearly more desirable and even indispensable when investigating $w$-modes or the gravitational echo of rotating neutron stars; we will, therefore, implement consistent initial data also for rotating configurations in the future, as well as implementing constraint damping mechanism \cite{2005CQGra..22.3767} in order to improve the long-term accuracy and reliability of our code (even though we are already able to evolve for several hundreds of milliseconds with no problem). However, at the current stage, we are not concerned by the arbitrariness in the initial data with regards to the accuracy of the observables in our simulation as the system continues to oscillate at its natural frequencies.

\subsection{Comparison to Published Values}
\label{sec:codetests_comparison}

While the present work is the first perturbative and comprehensive study of fluid modes of fast rotating neutron stars accounting for a dynamic spacetime, some sporadic values have been published during the past years. The largest set of $f$-mode frequencies has been calculated by Zink~\etal~\cite{2010PhRvD..81h4055Z} (henceforth ZKSS) who investigated the CFS-instability of the $f$-mode in rapidly rotating polytropic stars using non-linear simulations. They constructed two previously used sequences of neutron star models with two different polytropic indices, $\Gamma = 2$ (sequence S) and $\Gamma = 2.5$ (sequence C), and computed the frequencies of the $l=|m|=2$ and $l=|m|=3$ $f$-mode for those models.

\begin{figure}[htbp]
    \centering
    \includegraphics[width=8.6cm]{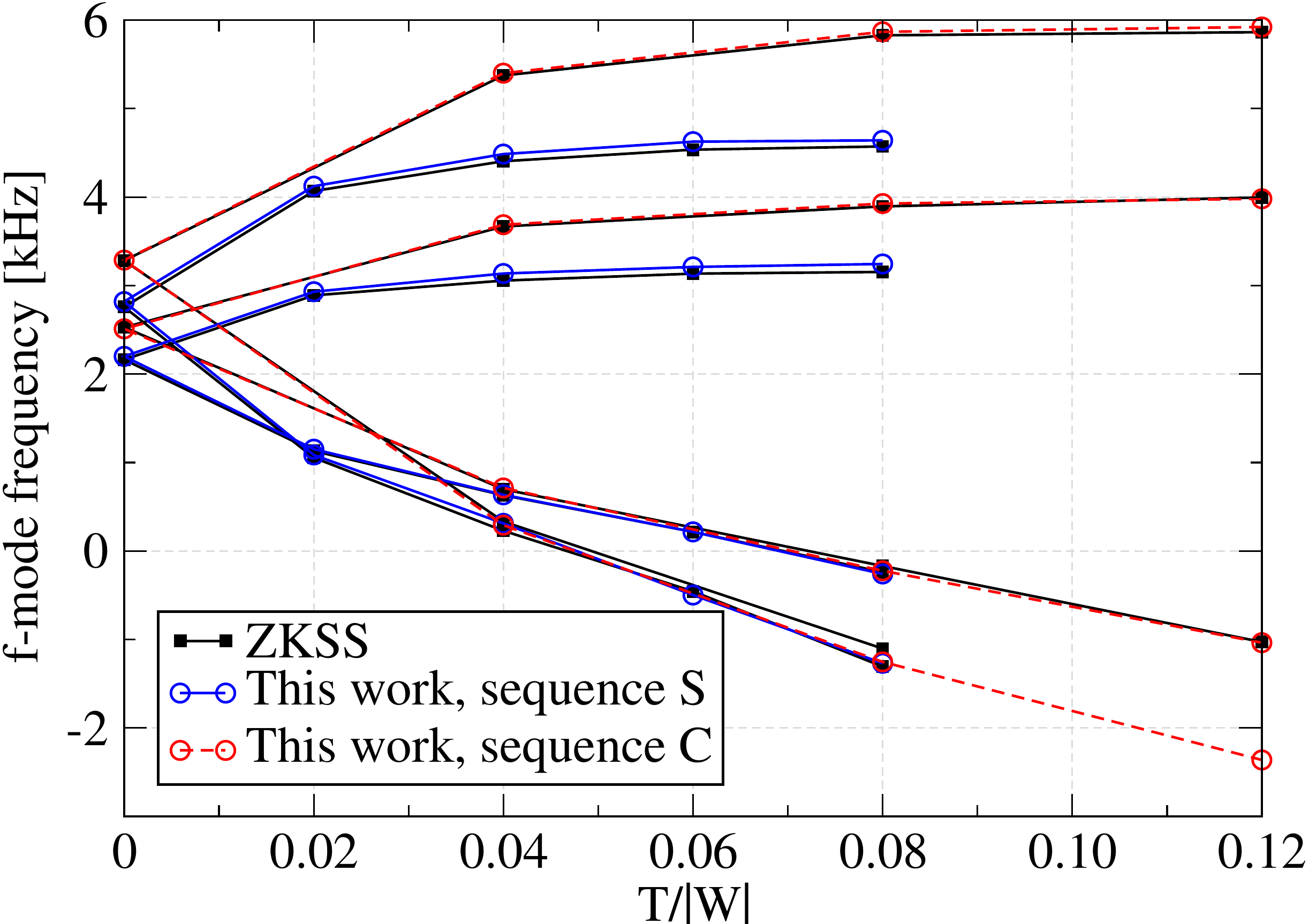}
    \caption{Graphical comparison of the frequencies reported by ZKSS (depicted by black squares) and those obtained by the present code (depicted by coloured circles). The excellent agreement of the two codes for various rotation rates (encoded in the ratio $T/|W|$) is obvious.}
    \label{fig:Zink}
\end{figure}

We constructed the same nine neutron star models as shown in Tables II and III in ZKSS,\footnote{Note that ZKSS introduce a separate length scale, $L_{\rm cgs}$, for their sequence C with which they scale their obtained $f$-mode frequencies; this length scale can be incorporated in the polytropic constant by setting $K = 2191$ which leads to modified values of the bulk properties of the equilibrium configurations. 
} and juxtaposed the $f$-mode frequencies obtained by our perturbative code and the published values in Tables~\ref{tab:Zink_l2} and \ref{tab:Zink_l3}. A graphical representation of the comparison is shown in Fig.~\ref{fig:Zink}. The frequencies show excellent agreement with a deviation of less than $2\,\%$ in the majority of cases; the relative difference between the two codes is naturally considerably larger for the potentially unstable $f$-mode where longer evolution times are necessary to achieve the same accuracy for the $f$-mode frequency.

\begin{table}[htbp]
    \centering
    \caption{Comparison of the frequencies of the $l=|m|=2$ $f$-mode obtained with the present code to the frequencies published by Zink~\etal~\cite{2010PhRvD..81h4055Z} for their sequences S and C. $\sigma({}^2f_2)$ is the frequency of the counter-rotating $f$-mode, $\sigma({}^2f_{-2})$ is the frequency of the co-rotating mode. All frequencies are given in kHz.}
    \begin{tabular}{ccd{3.3}d{3.3}d{3.3}d{3.3}}
        \toprule
        Model &  $T/|W|$ & \multicolumn{2}{c}{$\sigma({}^2f_2)$} & \multicolumn{2}{c}{$\sigma({}^2f_{-2})$} \\
        \cmidrule(lr){3-4}
        \cmidrule(lr){5-6}
              &  & \mc{ZKSS} & \mc{This work} & \mc{ZKSS} & \mc{This work} \\
        \midrule
        S0 & 0.00 &  2.162 &  2.199 & 2.162 & 2.199 \\
        S1 & 0.02 &  1.130 &  1.150 & 2.888 & 2.930 \\
        S2 & 0.04 &  0.630 &  0.635 & 3.056 & 3.135 \\
        S3 & 0.06 &  0.215 &  0.190 & 3.134 & 3.210 \\
        S4 & 0.08 & -0.240 & -0.260 & 3.154 & 3.244 \\
        \midrule
        C0 & 0.00 &  2.527 &  2.510 & 2.527 & 2.510 \\
        C1 & 0.04 &  0.698 &  0.712 & 3.668 & 3.687 \\
        C2 & 0.08 & -0.170 & -0.224 & 3.894 & 3.926 \\
        C3 & 0.12 & -1.029 & -1.037 & 3.995 & 3.980 \\
        \bottomrule
    \end{tabular}
    \label{tab:Zink_l2}
\end{table}

\begin{table}[htbp]
    \centering
    \caption{Same as Table~\ref{tab:Zink_l2} but for the $l=|m|=3$ $f$-mode.}
    \begin{tabular}{ccd{3.3}d{3.3}d{3.3}d{3.3}}
        \toprule
        Model &  $T/|W|$ & \multicolumn{2}{c}{$\sigma({}^3f_3)$} & \multicolumn{2}{c}{$\sigma({}^3f_{-3})$} \\
        \cmidrule(lr){3-4}
        \cmidrule(lr){5-6}
              &  & \mc{ZKSS} & \mc{This work} & \mc{ZKSS} & \mc{This work} \\
        \midrule
        S0 & 0.00 &  2.758 &  2.818 & 2.758 & 2.818 \\
        S1 & 0.02 &  1.054 &  1.085 & 4.067 & 4.123 \\
        S2 & 0.04 &  0.229 &  0.312 & 4.405 & 4.485 \\
        S3 & 0.06 & -0.457 & -0.498 & 4.535 & 4.626 \\
        S4 & 0.08 & -1.307 & -1.269 & 4.572 & 4.641 \\
        \midrule
        C0 & 0.00 &  3.282 &  3.287 & 3.282 & 3.287 \\
        C1 & 0.04 &  0.332 &  0.292 & 5.374 & 5.400 \\
        C2 & 0.08 & -1.104 & -1.254 & 5.828 & 5.867 \\
        C3 & 0.12 & \mc{$-$} & -2.362 & 5.864 & 5.921 \\
        \bottomrule
    \end{tabular}
    \label{tab:Zink_l3}
\end{table}

For completeness, we also report the $f$-mode frequencies of the non-rotating models as obtained by our eigenvalue code. The model S0 yields $f$-mode frequencies of $\sigma({}^2f) = 2.197\unit{kHz}$ and $\sigma({}^3f) = 2.811\unit{kHz}$, respectively, while for the model C0 we report $\sigma({}^2f) = 2.520\unit{kHz}$ and $\sigma({}^3f) = 3.291\unit{kHz}$. A comparison to the frequencies listed in Tables~\ref{tab:Zink_l2} and \ref{tab:Zink_l3} shows that all three codes produce values that are in strong agreement with each other.

Beside the work by ZKSS, we are aware of only one further data point at this time. Chaurasia~\etal~\cite{2018PhRvD..98j4005C} report the $f$-mode frequency extracted from full numerical relativity simulations of highly eccentric spinning binary neutron stars. Owing to the absence of results for $f$-mode frequencies in full general relativity at their time of publishing, they resorted to an intricate estimate for the $f$-mode frequency based on perturbation theory by starting from values and universal relations obtained in the Cowling approximation and applying rough corrections for effects of a dynamic spacetime. They ultimately arrive at a perturbation theory estimate for the frequency of the co-rotating $f$-mode of $2.20\unit{kHz}$, which they claim to be $\sim{}5\%$ larger than their observed frequency; thus, the $f$-mode frequency observed in their simulation should be $\sim 2.09\unit{kHz}$. For comparison, our proposed fitting formula, cf.~Eq.~(6) in \cite{2020PhRvL.125l1106K}, yields an estimate of $2.156\unit{kHz}$.

Following their analysis, we constructed a neutron star model using the tabulated EoS SLy that has a baryon mass of $M_0 = 1.504\,M_\odot$ and an angular rotation rate of $\Omega = 2\pi \cdot 191\unit{Hz}$; according to the \texttt{rns}-code, we find that this model has a central energy density of $\epsilon_c = 0.962 \cdot 10^{15} \unit{g/cm}^3$ and is rotating at $\Omega/\Omega_K = 0.166$ of its Kepler limit (where $\Omega_K$ is the angular velocity of a neutron star at the mass-shedding limit with the same central energy density). For our test neutron star, our evolution code yields the frequencies $1.655\unit{kHz}$ and $2.105\unit{kHz}$ for the counter- and co-rotating branch of the $f$-mode, respectively. At a relative difference of less than $1\%$, the results by Chaurasia~\etal~\cite{2018PhRvD..98j4005C} and our perturbative time evolution code are in excellent agreement.


\section{Summary and Outlook}

We present a set of time evolution equations governing the perturbations of the fluid of a rapidly rotating neutron star as well as its surrounding spacetime, derived in a perturbative framework in full general relativity. We have opted for the Hilbert gauge in order to arrive at a set of fully hyperbolic equations for the spacetime perturbations whose implementation does not pose major obstacles, while our treatment of the fluid is---with the exception for the correctional terms due to the spacetime perturbations---borrowed from previous studies on fast rotating neutrons stars.

For the numerical implementation, we have constructed a particular radial grid with a varying physical resolution; it resolves the surface of oblate neutron stars well, but becomes coarser with increasing radius, allowing it to extend far into the wave zone with a considerably less than proportional increase in computational expense. Like in previous studies, our hydrodynamical evolution is spoiled by numerical instabilities; we apply Kreiss-Oliger dissipation to damp out those spurious oscillations. We choose $3120 \times 50$ grid points as our ``standard resolution'' and on such a grid, the dissipation coefficients required for a stable evolution are of the order of $10^{-8}$ for the spacetime perturbations and $\approx 10^{-7} - 10^{-6}$ for the fluid perturbations. We sketch a proof that the Hilbert gauge (mathematically) remains satisfied throughout the evolution if it is satisfied in the initial data; in our simulations, we observe only very small numerical violations of the Hilbert gauge that remain bounded if we pick such initial data (even though we are not required to do so when concerned with fluid modes). We are planning to implement constraint damping mechanism for improved accuracy and also to construct initial data consistent with the Hilbert gauge for rotating configurations.

We perform convergence tests to study the accuracy of our code. With increasing grid resolution, we observe a clear convergence of the obtained frequencies toward a limit for both non-rotating and rotating background models. The convergence tests show that in a typical simulation, the obtained frequency usually deviates less than $1\% - 2\%$ from the limiting value or has an accuracy of about $10\unit{Hz}$. For non-rotating models, we can compare our obtained frequencies also with results from high-accuracy eigenvalue codes; the agreement between the results from our time evolution code and the values from the eigenvalue code is excellent. Furthermore, we reproduced previously published results from non-linear codes for several neutron star models at varying rotation rate with our perturbative code; again, the agreement between our values and the published ones is excellent for all considered models, irrespective of the rotation rate.

Our formulation of the problem allows to reduce the problem to two spatial dimensions which drastically lowers the computational expense of our numerical time evolutions in comparison to non-linear codes that perform three-dimensional simulations. The evolution of the perturbations of a neutron star for $15\unit{ms}$ on a grid with $3120 \times 50$ points, which facilitates a decent frequency resolution, requires only a few dozen of CPU hours; this enables us to study broad ranges of parameters and various EoSs. In the future, we expect to reduce the computational cost even further by deriving a simplified set of perturbation equations.

This concludes the presentation of the perturbation equations, their numerical implementation and accuracy tests of our code. Equipped with this code, we are able to produce a plethora of results concerning various oscillation modes of neutron stars; the most fundamental results concerning the fundamental mode of neutron stars are published in an accompanying paper~\cite{2020PhRvL.125l1106K}.

Beside the investigation of mode frequencies, we will also study the emission of gravitational waves associated with the various modes, in which the quadrupole formula seems the most suitable candidate as extraction method to complement our code.


\begin{acknowledgments}
We would like to thank the anonymous referee for constructive comments that helped us improve the manuscript. This work was supported by DFG research Grant No. 413873357. A part of the computations were performed on Trillian, a Cray XE6m-200 supercomputer at UNH supported by the NSF MRI program under Grant No. PHY-1229408.
\end{acknowledgments}

\begin{widetext}

\appendix

\section{Conventions and Abbreviations}
\label{app:abbreviations}

We introduce a few notational conventions and abbreviations that we use throughout all further appendices for the sake of brevity and clarity.

\begin{itemize}
    \item We use calligraphic letters for the spacetime perturbations and 
    \begin{equation}
        \Xi = \left\{ \mcA, \mcB, \mcH, \mcK, \mcL, \mcM, \mcP, \mcQ, \mcW, \mcY \right\}
    \end{equation}
    denotes the set of all ten spacetime perturbations.
    
    \item We use the comma notation to abbreviate partial derivatives, for example,
    \begin{equation}
        \mcK_{,t} := \frac{\partial}{\partial t} \mcK,
        \quad
        \omega_{,r} := \frac{\partial}{\partial r} \omega,
        \quad\text{ or }\quad
        \psi_{,r\theta} := \frac{\partial^2}{\partial\theta \partial r} \psi.
    \end{equation}
    
    \item We define the following shortcuts for some expressions:
    \begin{align}
        \kappa & := 4\pi \left( \epsilon + p \right),
        \label{eq:def_kappa} \\
        \kappa_1 & := 4\pi e^{2\nu} \left( u^t \right)^2 \left( \epsilon + p \right),
        \label{eq:def_kappa1} \\
        \kappa_2 & := \left( u^t \right)^2 (\epsilon + p),
        \label{eq:def_kappa2} \\
        \varpi & := \Omega - \omega.
        \label{eq:def_varpi}
    \end{align}

\end{itemize}


\section{Hilbert constraints}
\label{app:hilbert_constraints}

The definition of the Hilbert gauge, \cref{eq:hilbert_gauge}, can be expanded into four differential equations for the spacetime perturbations. Using the expressions \cref{eq:metricpert_tt,eq:metricpert_tr,eq:metricpert_tth,eq:metricpert_tph,eq:metricpert_rr,eq:metricpert_rth,eq:metricpert_rph,eq:metricpert_thth,eq:metricpert_thph,eq:metricpert_phph}, we find
\begin{align}
    e^{2\mu} \frac{\partial \mcH}{\partial t}
        & = - \frac{1}{2} e^{2\psi} \omega_{,r} r \sin\theta \mcA
            - \frac{1}{2} e^{2\psi} \omega_{,\theta} \sin\theta \mcB
            - im e^{2\mu} \omega \mcH
            - \frac{1}{2} \left( \psi_{,r} + \nu_{,r} + \frac{2}{r} \right) \mcL
            - \frac{1}{2} \mcL_{,r} \label{eq:hilbert_cn1} \\
        & \qquad
            - \frac{1}{2r} \left( \psi_{,\theta} + \nu_{,\theta} + \cot\theta \right) \mcM
            - \frac{1}{2r} \mcM_{,\theta}
            + \frac{im e^{2\nu+2\mu}}{2 r \sin\theta} \mcY, \nonumber \\
    e^{-2\nu} \frac{\partial \mcL}{\partial t}
        & = - \frac{im}{r\sin\theta} \mcA
            - 2 \nu_{,r} \mcH 
            + 2\left( \psi_{,r} + \nu_{,r} + \mu_{,r} + \frac{2}{r} \right) \mcK
            + 2 \mcK_{,r}
            - im e^{-2\nu} \omega \mcL 
            - 2 \left( \mu_{,r} + \frac{1}{r} \right) \mcP \label{eq:hilbert_cn2} \\
        & \qquad + \frac{1}{r} \left( \psi_{,\theta} + \nu_{,\theta} + 2\mu_{,\theta} + \cot\theta \right) \mcQ
            + \frac{1}{r} \mcQ_{,\theta}
            - 2 \left(\psi_{,r} + \frac{1}{r} \right) \mcW
            + e^{2\psi} \omega_{,r} r \sin\theta \mcY, \nonumber \\
    e^{-2\nu} \frac{\partial \mcM}{\partial t}
        & = - \frac{im}{r\sin\theta} \mcB
            - \frac{2 \nu_{,\theta}}{r} \mcH
            - \frac{2 \mu_{,\theta}}{r} \mcK
            - im e^{-2\nu} \omega \mcM
            + \frac{2}{r} \left( \psi_{,\theta} + \nu_{,\theta} + \mu_{,\theta} + \cot\theta \right) \mcP
            + \frac{2}{r} \mcP_{,\theta} \label{eq:hilbert_cn3} \\
        & \qquad + \left(\psi_{,r} + \nu_{,r} + 2\mu_{,r} + \frac{3}{r} \right) \mcQ
            + \mcQ_{,r}
            - \frac{2}{r} \left(\psi_{,\theta} + \cot\theta \right) \mcW
            + e^{2\psi} \omega_{,\theta} \sin\theta \mcY, \nonumber \\
    e^{2\mu} \frac{\partial \mcY}{\partial t}
        & = \left( 3\psi_{,r} + \nu_{,r} + \frac{3}{r} \right) \mcA
            + \mcA_{,r}
            + \frac{1}{r} \left( 3 \psi_{,\theta} + \nu_{,\theta} + 2 \cot\theta \right) \mcB
            + \frac{1}{r} \mcB_{,\theta}
            - \frac{2im e^{2\mu}}{e^{2\psi} r \sin\theta } \mcW
            - im e^{2\mu} \omega \mcY. \label{eq:hilbert_cn4} 
\end{align}


\section{Perturbations of the energy-momentum tensor}
\label{app:pert_emt}

In this appendix, we discuss the relations between the variables $Q_i$ that we defined in \cref{eq:def_Q1,eq:def_Q2,eq:def_Q3,eq:def_Q4,eq:def_Q5,eq:def_Q6}. It can easily be shown that $Q_5$ is, by definition, a linear combination of $Q_1$, $Q_2$, and $Q_6$,
\begin{align}
    Q_5
        & = - \Omega^2 r^2 \sin^2\theta Q_1 + 2\Omega r \sin\theta Q_2
        + g_{rr}^{(0)} \left( \Omega^2 g^{tt}_{(0)} - 2\Omega g^{t\varphi}_{(0)} + g^{\varphi\varphi}_{(0)} \right) r^2 \sin^2\theta Q_6 \\
        & = - \Omega^2 r^2 \sin^2\theta Q_1 + 2\Omega r \sin\theta Q_2
        + \left( \frac{e^{2\mu}}{e^{2\psi}} - e^{2\mu-2\nu} \varpi^2 r^2 \sin^2\theta \right) Q_6.
\end{align}
Furthermore, $Q_6$ is not an independent variable, too, but can be written as a linear combination of $Q_1$, $Q_2$, and some spacetime perturbations. To see this, we need to employ the definition of the speed of sound of the fluid,
\begin{align}
    \delta p & = c_s^2 \delta \epsilon,
\end{align}
and express the perturbations of pressure and energy density in terms of the $Q_i$'s and the spacetime perturbations. After some algebraic manipulations, we arrive at
\begin{align}
    Q_6
        & = \frac{e^{-2\mu+2\nu} c_s^2}{e^{2\nu} - e^{2\psi} r^2 \sin^2\theta \varpi^2 c_s^2} \times \\
        & \qquad \times \left[ \vphantom{\left(\Omega\right)^2}
                \left( e^{2\nu} + e^{2\psi} r^2 \sin^2\theta \left(\Omega^2 - \omega^2 \right) \right) Q_1
                - 2 e^{2\psi} r \sin\theta \varpi Q_2 \right. \nonumber \\
        & \qquad\qquad \left. \vphantom{\left(\Omega\right)^2}
                + 2 e^{2\nu} \kappa_2 \left( \mcH + e^{2\psi} r\sin\theta \varpi \mcY - e^{-2\nu+2\psi} r^2 \sin^2\theta \varpi^2 \mcW\right)
                - \left(\epsilon + p \right) \left( \mcH + \mcK + \mcP + \mcW \right)
            \right]. \nonumber
\end{align}
Compared to the corresponding expression in the Cowling approximation (see, e.g., Eq. (11) in \cite{2010PhRvD..81h4019K}), this expression has gained some correctional terms from the spacetime perturbations (and accounts for the slightly modified definition of $Q_2$) but is otherwise unaltered.


\section{Equations Governing the Spacetime Dynamics}
\label{app:spacetime}

In this appendix, we present the wave equations governing the dynamics of the ten spacetime perturbations. In order to simplify the expressions, we make liberal use of the background equations, i.e. those equations that emerge from the unperturbed Einstein equations, cf. Eq.~\eqref{eq:Einstein}, as well as the Hilbert conditions, cf. \cref{eq:hilbert_cn1,eq:hilbert_cn2,eq:hilbert_cn3,eq:hilbert_cn4}. 
The ten wave equations all take the same form and for any spacetime perturbation, say, $\mcX$ its corresponding wave equation can be written as
\begin{align}
    e^{2\mu-2\nu} \frac{\partial^2 \mcX}{\partial t^2}
        & = \sum_{\text{\Bicycle} \in \Sigma_1} \left[ \mcX | \text{\Bicycle} \right] \text{\Bicycle},
    \label{eq:wave_equations}
\end{align}
where we denote with the symbol $\left[ \mcX | \text{\Bicycle} \right]$ the coefficient of the function $\text{\Bicycle}$ appearing in the wave equation for $\mcX$, and $\Sigma_1$ is the set containing all ten spacetime perturbations, their first time derivatives, their first and second radial and polar derivatives, and the fluid perturbations, i.e.
\begin{align}
    \Sigma_1
        & := \left\{ Q_1, Q_2, Q_3, Q_4, Q_6 \right\}
        \cup \bigcup_{\mcX \in \Xi} \left\{ \mcX, \mcX_{,t}, \mcX_{,r}, \mcX_{,rr}, \mcX_{,\theta}, \mcX_{,\theta\theta} \right\},
\end{align}
where $\Xi$ is the set of all spacetime perturbation variables. In the following, we list all coefficients of the wave equations that are non-zero (and use the definitions shown in Appendix~\ref{app:abbreviations}):
\begingroup
\allowdisplaybreaks
\begin{align}
\left[ \mcA | \mcA \right]
    & = e^{2\mu} \left(2 \kappa - 4\kappa_1 - m^2 g_{(0)}^{\varphi\varphi} + 32\pi p \right) -\left(3 \psi_{,r}^2 - 2 \psi_{,r} \nu_{,r} + 6 \mu_{,r} \psi_{,r} - \nu_{,r}^2 + 2 \mu_{,r} \nu_{,r} \right) - e^{2\psi-2\nu}  \omega_{,\theta}^2 \sin^2\theta \\
    & \qquad -\frac {1}{r^2}\left[12 \psi_{,r} r + 6 \mu_{,r} r + 3 \psi_{,\theta}^2 +4 \psi_{,\theta} \nu_{,\theta} + \left( 9 \psi_{,\theta} + 5 \nu_{,\theta} - 2 \mu_{,\theta} + \cot\theta \right) \cot\theta + \nu_{,\theta}^2 +\nu_{,\theta\theta} +3\psi_{,\theta\theta} +5\right] \nonumber \\
\left[ \mcA | \mcA_{,t} \right]
    & = - 2 im e^{2\mu-2\nu} \omega \\
\left[ \mcA | \mcA_{,r} \right]
    & = 3 \psi_{,r} + 3 \nu_{,r} - 2 \mu_{,r} +\frac{2}{r}   \\
\left[ \mcA | \mcA_{,rr} \right]
    & = 1 \\
\left[ \mcA | \mcA_{,\theta} \right]
    & = \frac {1}{r^2} \left( - 2 \mu_{,\theta} + 3 \psi_{,\theta} +\nu_{,\theta} +\cot\theta \right)  \\
\left[ \mcA | \mcA_{,\theta\theta} \right]
    & = {\frac{1}{r^2}} \\
\left[ \mcA | \mcB \right]
    & = \frac{1}{r}\left(\nu_{,r} - \mu_{,r} - \frac{1}{r}\right) \left( 6 \psi_{,\theta} + 2 \nu_{,\theta} + 4 \cot\theta\right) + \frac{1}{r}\left(3 \psi_{,r\theta}+ \nu_{,r\theta}\right)
    + e^{2\psi-2\nu} \omega_{,r} \omega_{,\theta} r \sin^2\theta    \\
\left[ \mcA | \mcB_{,r} \right]
    & = \frac{2}{r} \mu_{,\theta} \\
\left[ \mcA | \mcB_{,\theta} \right]
    & = \frac {2}{r^2} \left( \nu_{,r} r - \mu_{,r} r - 1 \right) \\
\left[ \mcA | \mcK \right]   
    & = -2im e^{2\mu} \left( {\frac {2 \left( \psi_{,r} r+1 \right) }{e^{2\psi} r^2 \sin\theta }}+ e^{-2\nu}\omega_{,r} \omega r \sin\theta \right)  \\
\left[ \mcA | \mcK_{,t} \right]
    & = -2 e^{2\mu-2\nu} \omega_{,r} r\sin\theta \\
\left[ \mcA | \mcL \right]
    & = e^{-2\nu} \sin\theta \left( \psi_{,r}  r + \nu_{,r} r + 2  \right) \omega_{,r} + e^{-2\nu} \frac {1}{r} \sin\theta \left( 3 \psi_{,\theta} - \nu_{,\theta} -2 \mu_{,\theta} + 3 \cot\theta  +\partial_\theta \right) \omega_{,\theta} \\
\left[ \mcA | \mcL_{,r} \right]
    & = e^{-2\nu} \omega_{,r} r\sin\theta \\
\left[ \mcA | \mcL_{,\theta} \right]
    & = \frac{1}{r} e^{-2\nu} \omega_{,\theta} \sin\theta \\
\left[ \mcA | \mcM \right]
    & = - e^{-2\nu} \sin\theta \left( 2 \psi_{,\theta} - 2 \nu_{,\theta} - 2 \mu_{,\theta} + 2 \cot\theta +\partial_\theta \right) \omega_{,r} \\
\left[ \mcA | \mcQ \right]
    & = - im e^{2\mu} \left( {\frac { 2\left( \psi_{,\theta} +\cot\theta \right) }{ e^{2\psi} r^2 \sin\theta}}+ e^{-2\nu} \omega_{,\theta} \omega \sin\theta \right)   \\
\left[ \mcA | \mcQ_{,t} \right]
    & = - e^{2\mu-2\nu} \omega_{,\theta} \sin\theta \\
\left[ \mcA | \mcW \right]
    & = 2 im e^{2\mu} \left( {\frac {2 \left( \psi_{,r} r- \nu_{,r} r+1 \right) }{e^{2\psi} r^2 \sin\theta}}+ e^{-2\nu}\omega_{,r} \omega r \sin\theta \right)   \\
\left[ \mcA | \mcW_{,t} \right]
    & = 2 e^{2\mu-2\nu} \omega_{,r} r\sin\theta \\
\left[ \mcA | \mcY \right]
    & = - im e^{2\mu} \omega_{,r}  \\
\left[ \mcA | Q_{3} \right]
    & = -16 \pi e^{4\mu} \varpi r \sin\theta \\
\left[ \mcB | \mcA \right]
    & = \frac{1}{r^2} \left[ 2 \left( 3 \psi_{,r} r + \nu_{,r} r + 3 \right) \left(  \nu_{,\theta} - \mu_{,\theta} \right) - 2 \left(\mu_{,r} r + 1 \right) \cot\theta + 3 \psi_{,r\theta} r+ \nu_{,r\theta} r \right]+ e^{2\psi-2\nu} \omega_{,r} \omega_{,\theta} r \sin^2\theta \\
\left[ \mcB | \mcA_{,r} \right]
    & = \frac{2}{r} \left(  \nu_{,\theta} -\mu_{,\theta} \right) \\
\left[ \mcB | \mcA_{,\theta} \right]
    & = \frac{2}{r^2} \left( \mu_{,r} r+1 \right) \\
\left[ \mcB | \mcB \right]
    & =  - e^{2\mu} m^2 g_{(0)}^{\varphi\varphi} + e^{2\psi-2\nu}  \omega_{,\theta}^2 \sin^2\theta \\
    & \qquad + \frac{1}{r^2} \left[ \left( 2\nu_{,\theta} - 2\mu_{,\theta} - \cot\theta + \partial_\theta \right) \left( 3 \psi_{,\theta} + \nu_{,\theta} \right)   + 4 \left( \nu_{,\theta} - \mu_{,\theta} - \cot\theta\right) \cot\theta - 2 \right]    \nonumber \\
\left[ \mcB | \mcB_{,t} \right]
    & = - 2 im e^{2\mu-2\nu} \omega  \\
\left[ \mcB | \mcB_{,r} \right]
    & =  3 \psi_{,r} + \nu_{,r}  - 2 \mu_{,r} + \frac{2}{r} \\
\left[ \mcB | \mcB_{,rr} \right]
    & = 1 \\
\left[ \mcB | \mcB_{,\theta} \right]
    & = \frac {1}{r^2} \left( - 2 \mu_{,\theta} + 3 \nu_{,\theta} + 3 \psi_{,\theta} +\cot\theta \right) \\
\left[ \mcB | \mcB_{,\theta\theta} \right]
    & = \frac{1}{r^2} \\
\left[ \mcB | \mcL \right]
    & = - e^{-2\nu} \sin\theta \left( 2 \psi_{,r} - 2 \nu_{,r} - 2 \mu_{,r} + \partial_r \right) \omega_{,\theta} \\
\left[ \mcB | \mcM \right]
    & =  - 4 \kappa_1 e^{2\mu-2\nu}\varpi r\sin\theta - e^{-2\nu} \sin\theta \left( 2 \mu_{,r} r +1\right) \omega_{,r}- \frac{e^{-2\nu}}{r} \sin\theta \left(2 \psi_{,\theta} -2 \nu_{,\theta} + 2 \cot\theta  +\partial_\theta \right) \omega_{,\theta}    \\
\left[ \mcB | \mcM_{,r} \right]
    & = e^{-2\nu}  \omega_{,r} r\sin\theta  \\
\left[ \mcB | \mcM_{,\theta} \right]
    & = e^{-2\nu} \omega_{,\theta} r \sin\theta \\
\left[ \mcB | \mcP \right]
    & = -2im e^{2\mu} \left( \frac { 2 \left( \psi_{,\theta} +\cot\theta \right) }{e^{2\psi} r^2 \sin\theta} + e^{-2\nu} \omega_{,\theta} \omega \sin\theta \right)   \\
\left[ \mcB | \mcP_{,t} \right]
    & = -2 e^{2\mu-2\nu} \omega_{,\theta} \sin\theta  \\
\left[ \mcB | \mcQ \right]
    & = - im e^{2\mu} \left( {\frac {2 \left( \psi_{,r} r+1 \right) }{ e^{2\psi} r^2 \sin\theta}} +  e^{-2\nu} \omega_{,r} \omega r \sin\theta \right)  \\
\left[ \mcB | \mcQ_{,t} \right]
    & = - e^{2\mu-2\nu} \omega_{,r} r\sin\theta \\
\left[ \mcB | \mcW \right]
    & = 2 im e^{2\mu}\left( \frac {2 \left( \psi_{,\theta} -\nu_{,\theta} +\cot\theta \right) }{ e^{2\psi} r^2 \sin\theta} + e^{-2\nu}  \omega_{,\theta} \omega \sin\theta \right)   \\
\left[ \mcB | \mcW_{,t} \right]
    & = 2 e^{2\mu-2\nu} \omega_{,\theta} \sin\theta  \\
\left[ \mcB | \mcY \right]
    & = -\frac { e^{2\mu}}{r} im \omega_{,\theta} \\
\left[ \mcB | Q_4 \right]
    & = -16 \pi e^{4\mu} \varpi r\sin\theta  \\
\left[ \mcH | \mcA \right]
    & =  \frac{2 im \nu_{,r}}{r\sin\theta}+ im e^{2\psi-2\nu} \omega \omega_{,r} r\sin\theta   \\
\left[ \mcH | \mcB \right]
    & = \frac {2 im \nu_{,\theta}}{r^2 \sin\theta} + im e^{2\psi-2\nu} \omega_{,\theta} \omega \sin\theta   \\
\left[ \mcH | \mcH \right]
    & =  e^{2\mu} \left( -2 m^2 e^{-2\nu} \omega^2 - \kappa + 2 \kappa_1 -{m}^{2}g_{(0)}^{\varphi\varphi} \right) + 2 \left( \nu_{,r}^2 + \frac{1}{r^2} \nu_{,\theta}^2 \right) + e^{2\psi-2\nu} \sin^2\theta  \left( r^2 \omega_{,r}^2 + \omega_{,\theta}^2  \right)    \\
\left[ \mcH | \mcH_{,r} \right]
    & =  \psi_{,r} + \nu_{,r} +\frac{2}{r}  \\
\left[ \mcH | \mcH_{,rr} \right]
    & = 1 \\
\left[ \mcH | \mcH_{,\theta} \right]
    & = \frac{1}{r^2} \left( \psi_{,\theta} +\nu_{,\theta} +\cot\theta \right) \\
\left[ \mcH | \mcH_{,\theta\theta} \right]
    & = \frac{1}{r^2} \\
\left[ \mcH | \mcK \right]
    & =  e^{2\mu} \left( \kappa - 16 \pi p \right) - e^{2\psi-2\nu} \omega_{,\theta}^2 \sin^2\theta -2 \left(\psi_{,r} + \nu_{,r} + \mu_{,r} + \frac{2}{r} \right) \nu_{,r} \\
    & \qquad
    + \frac{2}{r^2} \left(\psi_{,\theta} + \nu_{,\theta} + \mu_{,\theta} +  \cot\theta +\partial_\theta \right) \nu_{,\theta}  \nonumber \\
\left[ \mcH | \mcK_{,r} \right]
    & = -4 \nu_{,r}   \\
\left[ \mcH | \mcL \right]
    & = im e^{-2\nu} \left( \psi_{,r} + \nu_{,r} +\partial_r +\frac{2}{r} \right) \omega  \\
\left[ \mcH | \mcL_{,r} \right]
    & = i\omega m e^{-2\nu} \\
\left[ \mcH | \mcM \right]
    & = \frac{e^{-2\nu}}{r} im \left( \psi_{,\theta} +\nu_{,\theta} + \cot\theta +\partial_\theta \right) \omega \\
\left[ \mcH | \mcM_{,\theta} \right]
    & = \frac {1}{r}i\omega m e^{-2\nu} \\
\left[ \mcH | \mcP \right]
    & = \left( 4 \kappa_1 -\kappa \right) e^{2\mu} + 2 \left( \mu_{,r}+ \frac{1}{r} \right) \nu_{,r} - \frac{2}{r^2}\left( 2 \psi_{,\theta}  + 2 \nu_{,\theta} + \mu_{,\theta} + 2 \cot\theta +\partial_\theta \right) \nu_{,\theta} + e^{2\psi-2\nu} \omega_{,\theta}^2 \sin^2\theta    \\
\left[ \mcH | \mcP_{,\theta} \right]
    & = - \frac{4}{r^2} \nu_{,\theta} \\
\left[ \mcH | \mcQ \right]
    & =  -\frac{2}{r}  \left(\psi_{,r} + \mu_{,r} + \frac{2}{r} + \partial_r\right) \nu_{,\theta} - \frac{2}{r} \left( \psi_{,\theta} + 2 \nu_{,\theta} + \mu_{,\theta} + \cot\theta \right) \nu_{,r} + e^{2\psi-2\nu} \omega_{,r} \omega_{,\theta} r \sin^2\theta   \\
\left[ \mcH | \mcQ_{,r} \right]
    & = - \frac { 2 }{r} \nu_{,\theta} \\
\left[ \mcH | \mcQ_{,\theta} \right]
    & = - \frac { 2 }{r} \nu_{,r} \\
\left[ \mcH | \mcW \right]
    & =  e^{2\mu} \left( \kappa + 2 \kappa_1 \right) +2 \left( \psi_{,r} + \frac{1}{r} \right) \nu_{,r}
    +\frac{2}{r^2} \left( \psi_{,\theta} + \cot\theta \right) \nu_{,\theta} - e^{2\psi-2\nu} \sin^2\theta  \left( r^2 \omega_{,r}^2 + \omega_{,\theta}^2 \right)     \\
\left[ \mcH | \mcY \right]
    & =   e^{2\mu} \left( {\frac {{m}^{2}\omega }{r\sin\theta }} + 2 e^{2\psi}\varpi \kappa_1 r\sin\theta\right) - e^{2\psi} \sin\theta \left( 3 \nu_{,r} r - 1 \right) \omega_{,r} - \frac{e^{2\psi}}{r} \sin\theta \left(3 \nu_{,\theta} - \cot\theta \right) \omega_{,\theta}   \\
\left[ \mcH | \mcY_{,r} \right]
    & = - e^{2\psi} \omega_{,r} r \sin\theta  \\
\left[ \mcH | \mcY_{,\theta} \right]
    & = -{\frac { e^{2\psi} \omega_{,\theta} \sin\theta }{r}} \\
\left[ \mcH | Q_1 \right]
    & = -8\pi e^{2\mu+2\nu} \\
\left[ \mcK | \mcA \right]
    & =  im \left( \frac {2 \left( \psi_{,r} r- \nu_{,r} r+1 \right) }{r^2 \sin\theta} +  e^{2\psi-2\nu}\omega \omega_{,r} r\sin\theta \right)   \\
\left[ \mcK | \mcA_{,t} \right]
    & =  e^{2\psi-2\nu}  \omega_{,r} r \sin\theta   \\
\left[ \mcK | \mcH \right]
    & = e^{2\mu} \left( \kappa-2 \kappa_1 -16\pi p \right) - e^{2\psi-2\nu}  \omega_{,\theta}^2 \sin^2\theta
    + 2 \left( \psi_{,r} - \nu_{,r} + \mu_{,r} + \frac{2}{r} \right) \nu_{,r} \\
    & \qquad
        + \frac{2}{r^2}\left( \psi_{,\theta}  + \nu_{,\theta} - \mu_{,\theta} + \cot\theta +\partial_\theta \right) \nu_{,\theta}  \nonumber \\
\left[ \mcK | \mcK \right]
    & = - e^{2\mu} \left( \kappa+m^2 g_{(0)}^{\varphi\varphi} \right) + e^{2\psi-2\nu} \omega_{,r}^2 r^2 \sin^2\theta
    - 2\left( \psi_{,r}^2 - 2 \psi_{,r} \nu_{,r} - \nu_{,r}^2 -2\mu_{,r} \nu_{,r} + \mu_{,r}^2 \right) \\
    & \qquad  - \frac{4}{r} \left[ \psi_{,r} - 2 \nu_{,r} + \mu_{,r}  \right] - \frac{2}{r^2} \left[ \mu_{,\theta} ^{2}+2 \right] \nonumber \\
\left[ \mcK | \mcK_{,t} \right]
    & = - 2 im e^{2\mu-2\nu}\omega \\
\left[ \mcK | \mcK_{,r} \right]
    & = \psi_{,r} + 5 \nu_{,r} +\frac{2}{r} \\
\left[ \mcK | \mcK_{,rr} \right]
    & = 1 \\
\left[ \mcK | \mcK_{,\theta} \right]
    & = \frac{1}{r^2} \left( \psi_{,\theta} +\nu_{,\theta} +\cot\theta \right)  \\
\left[ \mcK | \mcK_{,\theta\theta} \right]
    & = \frac{1}{r^2} \\
\left[ \mcK | \mcL \right]
    & = -im e^{-2\nu} \omega_{,r}  \\
\left[ \mcK | \mcP \right]
    & =  e^{2\mu}\kappa - \frac{1}{2} e^{2\psi-2\nu} \left( r^2 \omega_{,r}^2 + \omega_{,\theta}^2  \right) \sin^2\theta
    - 2 \left[ \psi_{,r} \nu_{,r} + 2 \mu_{,r} \nu_{,r} - \mu_{,r}^2 \right]
    - \frac{2}{r} \left[ 3 \nu_{,r} - 2 \mu_{,r} \right]
    \\
    & \qquad
    - \frac{2}{r^2} \left[ \psi_{,\theta} \nu_{,\theta} + \nu_{,\theta} \cot\theta - \mu_{,\theta}^2 - 1 \right] \nonumber \\
\left[ \mcK | \mcQ \right]
    & = \frac{1}{r^2} \left( 2 \nu_{,r} r - 2 \mu_{,r} r - 2 \right) \left( \psi_{,\theta} + \nu_{,\theta}  + \cot\theta \right) + \frac{1}{r} \left( 4 \mu_{,\theta} \nu_{,r} + \psi_{,r\theta} + \nu_{,r\theta}  \right) \\
\left[ \mcK | \mcQ_{,r} \right]
    & = \frac { 2 }{r} \mu_{,\theta}\\
\left[ \mcK | \mcQ_{,\theta} \right]
    & = \frac{2}{r^2} \left( \nu_{,r} r- \mu_{,r} r-1 \right) \\
\left[ \mcK | \mcW \right]
    & = e^{2\mu} \left( 2 \kappa_1 -\kappa \right)  + \frac{1}{2} e^{2\psi-2\nu}  \left(3 \omega_{,\theta}^2  -r^2 \omega_{,r}^2  \right) \sin^2\theta
    + 2 \left[ \psi_{,r}^2 - 2 \psi_{,r} \nu_{,r} - \mu_{,r} \nu_{,r} \right]
    \\
    & \qquad
    + \frac{2}{r} \left[ 2 \psi_{,r} - 3 \nu_{,r} \right] - \frac{2}{r^2} \left[ \nu_{,\theta}^2 - \nu_{,\theta} \mu_{,\theta} +\nu_{,\theta\theta} -1 \right] \nonumber \\
\left[ \mcK | \mcY \right]
    & = -2\kappa_1 e^{2\psi+2\mu} \varpi r\sin\theta - e^{2\psi} \left( \psi_{,r} - \nu_{,r} + \mu_{,r} + \frac{2}{r} \right) \omega_{,r} r \sin\theta \\
    & \qquad
    - \frac{e^{2\psi}}{r} \sin\theta \left( 3 \psi_{,\theta} - \nu_{,\theta} - \mu_{,\theta} +3 \cot\theta  +\partial_\theta \right) \omega_{,\theta} \nonumber \\
\left[ \mcK | Q_6 \right]
    & = 8\pi e^{4\mu} \\
\left[ \mcL | \mcA \right]
    & = - e^{2\mu}
        \left( 4 \kappa_1 e^{2\psi}\varpi r\sin\theta  + 2 m^2 \frac {\omega }{r\sin\theta}
        \right)
        + 2 e^{2\psi} r\sin\theta \left( \psi_{,r} + \nu_{,r} - \mu_{,r} \right) \omega_{,r}
        \\
    & \qquad
    - \frac {e^{2\psi} \sin\theta}{r}
           \left( 3 \psi_{,\theta} - \nu_{,\theta} +4 \cot\theta +\partial_\theta \right) \omega_{,\theta} \nonumber \\
\left[ \mcL | \mcA_{,r} \right]
    & = 2 e^{2\psi} \omega_{,r} r \sin\theta   \\
\left[ \mcL | \mcA_{,\theta} \right]
    & = \frac { e^{2\psi}  \omega_{,\theta} \sin\theta}{r} \\
\left[ \mcL | \mcB \right]
    & = e^{2\psi} \sin\theta \left( 3 \psi_{,\theta} + \nu_{,\theta} +2 \cot\theta +\partial_\theta \right) \omega_{,r}
        + 2 e^{2\psi} \sin\theta \left( \psi_{,r} - \mu_{,r} \right) \omega_{,\theta} \\
\left[ \mcL | \mcB_{,\theta} \right]
    & = e^{2\psi} \omega_{,r} \sin\theta   \\
\left[ \mcL | \mcH \right]
    & = 2 im e^{2\mu} \left( 2 \omega \nu_{,r} +\omega_{,r} \right) \\
\left[ \mcL | \mcK \right]
    & = -2 im e^{2\mu} \left( 2 \psi_{,r} +2 \mu_{,r} +\partial_r +\frac{4}{r} \right) \omega  \\
\left[ \mcL | \mcK_{,t} \right]
    & = 4 e^{2\mu} \nu_{,r} \\
\left[ \mcL | \mcK_{,r} \right]
    & = -4 im e^{2\mu}\omega   \\
\left[ \mcL | \mcL \right]
    & = - e^{2\mu} \left(
        2 m^2 e^{-2\nu} \omega^2 + 16 \pi p + m^2 g_{(0)}^{\varphi\varphi}
        \right) - \left( \psi_{,r} - \nu_{,r} \right)^2 + 2 \mu_{,r} \left( \psi_{,r} + \nu_{,r} \right) 
        + e^{2\psi-2\nu} \omega_{,r}^2 r^2 \sin^2\theta
        \\
    & +\frac {1}{r^2} \left[ \psi_{,r} r + 5 \nu_{,r} r + 3 \psi_{,\theta}^2 - 4 \mu_{,\theta} \left( \psi_{,\theta} + \cot\theta \right) + 6 \psi_{,\theta} \cot\theta + \nu_{,\theta}^2 +\nu_{,\theta\theta} + 3 \psi_{,\theta\theta} -2\right]  \nonumber \\
\left[ \mcL | \mcL_{,r} \right]
    & = \psi_{,r} + \nu_{,r} - 2 \mu_{,r} +\frac{2}{r} \\
\left[ \mcL | \mcL_{,rr} \right]
    & = 1 \\
\left[ \mcL | \mcL_{,\theta} \right]
    & = \frac{1}{r^2} \left( \psi_{,\theta}  -\nu_{,\theta} - 2 \mu_{,\theta} +\cot\theta\right) \\
\left[ \mcL | \mcL_{,\theta\theta} \right]
    & = \frac{1}{r^2} \\
\left[ \mcL | \mcM \right]
    & = - \frac{1}{r}\left(2 \psi_{,\theta} - 2 \mu_{,\theta} + 2 \cot\theta + 
    \partial_\theta \right) \left(\psi_{,r}-\nu_{,r}\right) - \frac{1}{r^2}
    \left(2 \psi_{,\theta} - 4 \mu_{,\theta} + 2\cot\theta \right) \\
\left[ \mcL | \mcM_{,r} \right]
    & = \frac { 2 }{r} \mu_{,\theta} \\
\left[ \mcL | \mcM_{,\theta} \right]
    & = \frac{2}{r^2} \left( \nu_{,r} r- \mu_{,r} r-1 \right)  \\
\left[ \mcL | \mcP \right]
    & = 4 e^{2\mu} im\omega \left( \mu_{,r}+\frac{1}{r} \right) \\
\left[ \mcL | \mcQ \right]
    & = -\frac {im}{r} e^{2\mu}  \left( 2 \psi_{,\theta} + 4 \mu_{,\theta} + 2 \cot\theta + \partial_\theta \right) \omega \\
\left[ \mcL | \mcQ_{,t} \right]
    & = \frac{2}{r} e^{2\mu}\nu_{,\theta} \\
\left[ \mcL | \mcQ_{,\theta} \right]
    & = -\frac {2 im e^{2\mu}\omega  }{r} \\
\left[ \mcL | \mcW \right]
    & = 2 im e^{2\mu} \left( 2 \psi_{,r} \omega -\omega_{,r} + \frac{2}{r} \omega \right)   \\
\left[ \mcL | \mcY \right]
    & = 2 im e^{2\mu}
        \left(
            \frac { e^{2\nu} \left( \psi_{,r} r- \nu_{,r} r+1 \right) }{r^2 \sin\theta}
            - e^{2\psi}\omega \omega_{,r} r\sin\theta
        \right)  \\
\left[ \mcL | Q_3 \right]
    & = -16\pi e^{2\nu+4\mu} \\
\left[ \mcM | \mcA \right]
    & = e^{2\psi} \sin\theta \left( 3 \psi_{,r} + \nu_{,r} +\frac{3}{r} \right) \omega_{,\theta}
        +e^{2\psi} \sin\theta \left( 2 \psi_{,\theta} - 2 \mu_{,\theta} + 2 \cot\theta +\partial_\theta \right) \omega_{,r} \\
\left[ \mcM | \mcA_{,r} \right]
    & = e^{2\psi} \omega_{,\theta} \sin\theta  \\
\left[ \mcM | \mcB \right]
    & = -\frac{2 m^2 e^{2\mu}\omega }{r\sin\theta }+\frac{e^{2\psi}}{r} \sin\theta \left( 5 \psi_{,\theta} + \nu_{,\theta} - 2 \mu_{,\theta} + 3 \cot\theta +\partial_\theta \right) \omega_{,\theta}   \\
\left[ \mcM | \mcB_{,r} \right]
    & = e^{2\psi} \omega_{,r} r \sin\theta  \\
\left[ \mcM | \mcB_{,\theta} \right]
    & = \frac{2}{r} e^{2\psi} \omega_{,\theta} \sin\theta \\
\left[ \mcM | \mcH \right]
    & = \frac{2}{r} im e^{2\mu} \left( 2 \omega \nu_{,\theta} +\omega_{,\theta} \right) \\
\left[ \mcM | \mcK \right]
    & = \frac{4}{r} im e^{2\mu}\omega  \mu_{,\theta} \\
\left[ \mcM | \mcL \right]
    & = - e^{2\psi-2\nu} \omega_{,r} \omega_{,\theta} r \sin^2\theta + \frac{2}{r} \left( \nu_{,\theta} - \mu_{,\theta} \right) \left( \psi_{,r} + \nu_{,r} \right)  + \frac{1}{r} \left( \psi_{,r\theta} + 3 \nu_{,r\theta} - 4 \nu_{,\theta} \mu_{,r} \right) - \frac{4}{r^2} \mu_{,\theta}     \\
\left[ \mcM | \mcL_{,r} \right]
    & = \frac{2}{r} \left( \nu_{,\theta} -\mu_{,\theta} \right) \\
\left[ \mcM | \mcL_{,\theta} \right]
    & = \frac{2}{r^2} \left( \mu_{,r} r+1 \right) \\
\left[ \mcM | \mcM \right]
    & = e^{2\mu} \left(2 \kappa - 4 \kappa_1 - m^2 g_{(0)}^{\varphi\varphi} - 2 m^2 e^{-2\nu} \omega^2 \right)
        - \frac{1}{2} e^{2\psi-2\nu} \left( r^2 \omega_{,r}^2 + \omega_{,\theta}^2  \right) \sin^2\theta
        - \frac{1}{r} \left( \psi_{,r} + \nu_{,r} + 4 \mu_{,r} \right)
        \\
    & \qquad 
        + 2 \left( \mu_{,r} \nu_{,r} - \psi_{,r} \nu_{,r} - \mu_{,r} \psi_{,r} \right)
    + \frac{1}{r^2} \left( - 2 \psi_{,\theta}^2 - 4 \psi_{,\theta} \cot\theta - \cot^2\theta +\nu_{,\theta\theta} -\psi_{,\theta\theta} - 1 \right) \nonumber \\
\left[ \mcM | \mcM_{,r} \right]
    & = \psi_{,r} - \nu_{,r} - 2 \mu_{,r} + \frac{2}{r}  \\
\left[ \mcM | \mcM_{,rr} \right]
    & = 1 \\
\left[ \mcM | \mcM_{,\theta} \right]
    & = \frac {1}{r^2} \left(- 2 \mu_{,\theta} + \nu_{,\theta} + \psi_{,\theta} + \cot\theta \right)  \\
\left[ \mcM | \mcM_{,\theta\theta} \right]
    & = \frac{1}{r^2} \\
\left[ \mcM | \mcP \right]
    & = -\frac {2}{r} im e^{2\mu} \left( 2 \psi_{,\theta} + 2 \mu_{,\theta} + 2 \cot\theta + \partial_\theta \right) \omega \\
\left[ \mcM | \mcP_{,t} \right]
    & = \frac{4}{r} e^{2\mu} \nu_{,\theta} \\
\left[ \mcM | \mcP_{,\theta} \right]
    & = - \frac{4}{r} im e^{2\mu}\omega  \\
\left[ \mcM | \mcQ \right]
    & = - im  e^{2\mu}\left( 2 \psi_{,r} + 4 \mu_{,r} +\frac{6}{r} + \partial_r \right) \omega \\
\left[ \mcM | \mcQ_{,t} \right]
    & = 2 e^{2\mu} \nu_{,r}  \\
\left[ \mcM | \mcQ_{,r} \right]
    & = - 2im e^{2\mu}\omega   \\
\left[ \mcM | \mcW \right]
    & = \frac {2}{r} im e^{2\mu} \left( 2 \psi_{,\theta} + 2 \cot\theta - \partial_\theta \right) \omega \\
\left[ \mcM | \mcY \right]
    & = 2 im e^{2\mu} \left( {\frac { e^{2\nu} \left( \psi_{,\theta} -\nu_{,\theta} +\cot\theta \right) }{r^2 \sin\theta}} -  e^{2\psi}\omega\omega_{,\theta} \sin\theta \right)  \\
\left[ \mcM | Q_{4} \right]
    & = - 16\pi e^{4\mu+2\nu} \\
\left[ \mcP | \mcB \right]
    & = im \left( \frac {2 \left( \psi_{,\theta} -\nu_{,\theta} +\cot\theta \right) }{r^2 \sin\theta} + e^{2\psi-2\nu} \omega_{,\theta} \omega \sin\theta  \right)   \\
\left[ \mcP | \mcB_{,t} \right]
    & =  e^{2\psi-2\nu} \omega_{,\theta} \sin\theta  \\
\left[ \mcP | \mcH \right]
    & = e^{2\mu} \left( 2 \kappa_1 - \kappa\right) - 2\left( \mu_{,r} + \frac{1}{r} \right) \nu_{,r} - \frac{2}{r^2}\left( 2 \nu_{,\theta} - \mu_{,\theta} +\partial_\theta \right) \nu_{,\theta} + e^{2\psi-2\nu} \omega_{,\theta}^2 \sin^2\theta   \\
\left[ \mcP | \mcK \right]
    & = e^{2\mu}\kappa
    - \frac{1}{2} e^{2\psi-2\nu} \left( r^2 \omega_{,r}^2 + \omega_{,\theta}^2  \right) \sin^2\theta 
    - 2\left( \psi_{,r} + \frac{1}{r} \right) \nu_{,r}
    + 2\left( \mu_{,r} + \frac{2}{r} \right) \mu_{,r}
     \\
    & \qquad
    - \frac{2}{r^2}\left[ \nu_{,\theta} \left( \psi_{,\theta} +2 \mu_{,\theta} + \cot\theta \right) - \mu_{,\theta} ^{2}-1 \right]
    \nonumber  \\
\left[ \mcP | \mcM \right]
    & = - \frac{im}{r} e^{-2\nu} \omega_{,\theta}   \\
\left[ \mcP | \mcP \right]
    & = - e^{2\mu} \left( m^2 g_{(0)}^{\varphi\varphi} + \kappa \right)
    + e^{2\psi-2\nu} \omega_{,\theta}^2 \sin^2\theta
    - 2 \left( \mu_{,r} + \frac{2}{r} \right)\mu_{,r}
    - \frac{2}{r^2} \left[ \left( 2 \psi_{,\theta} - 2 \nu_{,\theta} + \cot\theta \right) \cot\theta +1 \right] \\
    & \qquad 
    - \frac{2}{r^2} \left( \psi_{,\theta}^2 - 2 \psi_{,\theta} \nu_{,\theta} - \nu_{,\theta}^2 -2 \nu_{,\theta} \mu_{,\theta}+ \mu_{,\theta}^2 \right)    
     \nonumber \\
\left[ \mcP | \mcP_{,t} \right]
    & = - 2 im e^{2\mu-2\nu}\omega  \\
\left[ \mcP | \mcP_{,r} \right]
    & = \psi_{,r} + \nu_{,r} + \frac{2}{r} \\
\left[ \mcP | \mcP_{,rr} \right]
    & = 1 \\
\left[ \mcP | \mcP_{,\theta} \right]
    & = \frac{1}{r^2} \left( \psi_{,\theta} + 5 \nu_{,\theta} + \cot\theta \right) \\
\left[ \mcP | \mcP_{,\theta\theta} \right]
    & = \frac{1}{r^2} \\
\left[ \mcP | \mcQ \right]
    & = \frac{1}{r^2} \left[ \left( \nu_{,\theta} - \mu_{,\theta} \right) \left( 2 \psi_{,r} r + 2 \nu_{,r} r + 2\right)  + 4 \nu_{,\theta} \left( \mu_{,r} r + 1 \right) + \psi_{,r\theta} r+ \nu_{,r\theta} r \right] \\
\left[ \mcP | \mcQ_{,r} \right]
    & = \frac{2}{r} \left( \nu_{,\theta} -\mu_{,\theta} \right)  \\
\left[ \mcP | \mcQ_{,\theta} \right]
    & = \frac{2}{r^2} \left( \mu_{,r} r+1 \right) \\
\left[ \mcP | \mcW \right]
    & = e^{2\mu} \left( \kappa - 2 \kappa_1 \right) - e^{2\psi-2\nu} \omega_{,\theta}^2 \sin^2\theta
    - \frac{2}{r^2} \left[ \left( 2 \nu_{,\theta} -  \mu_{,\theta} \right) \left( \psi_{,\theta} + \cot\theta \right) - \cot^2\theta +\psi_{,\theta\theta} \right]
    \\
    & \qquad
    - 2 \mu_{,r} \psi_{,r} - \frac{2}{r} \left( \psi_{,r} + \mu_{,r} \right)
    \nonumber\\
\left[ \mcP | \mcY \right]
    & = 2 \kappa_1 e^{2\psi+2\mu} \varpi r\sin\theta + e^{2\psi} \sin\theta \left( \mu_{,r} r + 1 \right) \omega_{,r} + \frac{e^{2\psi}}{r} \sin\theta \left(2 \psi_{,\theta} - \mu_{,\theta} +2 \cot\theta + \partial_\theta \right) \omega_{,\theta} \\
\left[ \mcP | Q_6 \right]
    & = 8\pi e^{4\mu} \\
\left[ \mcQ | \mcA \right]
    & = im \left( {\frac {2 \left( \psi_{,\theta} -\nu_{,\theta} +\cot\theta \right) }{r^2 \sin\theta}} + e^{2\psi-2\nu} \omega_{,\theta} \omega \sin\theta \right)   \\
\left[ \mcQ | \mcA_{,t} \right]
    & = e^{2\psi-2\nu} \omega_{,\theta} \sin\theta \\
\left[ \mcQ | \mcB \right]
    & = im \left( {\frac {2 \left( \psi_{,r} r- \nu_{,r} r+1 \right) }{r^2 \sin\theta}} + e^{2\psi-2\nu}\omega_{,r} \omega r\sin\theta \right)   \\
\left[ \mcQ | \mcB_{,t} \right]
    & =  e^{2\psi-2\nu} \omega_{,r} r \sin\theta \\
\left[ \mcQ | \mcH \right]
    & = 2 e^{2\psi-2\nu} \omega_{,r} \omega_{,\theta} r \sin^2\theta - \frac{4}{r^2}  \left[2 \nu_{,r} \nu_{,\theta} r- \mu_{,\theta} \nu_{,r} r- \nu_{,\theta} \mu_{,r} r+ \nu_{,r\theta} r-\nu_{,\theta} \right]    \\
\left[ \mcQ | \mcK \right]
    & = \frac{2}{r^2} \left[ \left( 2 \nu_{,\theta} - 2 \mu_{,\theta} \right) \left(\psi_{,r} r +1 \right) + 2 \nu_{,\theta} \left( \nu_{,r} r + \mu_{,r} r + 1\right) - 4 \mu_{,\theta} \nu_{,r} r+ \psi_{,r\theta} r+ \nu_{,r\theta} r \right] \\
\left[ \mcQ | \mcK_{,r} \right]
    & = \frac{4}{r} \left( \nu_{,\theta} -\mu_{,\theta} \right) \\
\left[ \mcQ | \mcK_{,\theta} \right]
    & = \frac{4}{r^2}\left( \mu_{,r} r+1 \right) \\
\left[ \mcQ | \mcL \right]
    & = - \frac{im}{r} e^{-2\nu} \omega_{,\theta} \\
\left[ \mcQ | \mcM \right]
    & = -im e^{-2\nu} \omega_{,r} \\
\left[ \mcQ | \mcP \right]
    & = \frac{2}{r^2} \left[ \left( 2 \nu_{,r} r - 2 \mu_{,r} r - 2\right) \left( \psi_{,\theta} + 2\nu_{,\theta} +\cot\theta \right) + 2 \nu_{,r} r \left( \mu_{,\theta} - \nu_{,\theta} \right) + \psi_{,r\theta} r+ \nu_{,r\theta} r \right] \\
\left[ \mcQ | \mcP_{,r} \right]
    & = \frac{4}{r} \mu_{,\theta}  \\
\left[ \mcQ | \mcP_{,\theta} \right]
    & = \frac{4}{r^2}\left( \nu_{,r} r- \mu_{,r} r-1 \right) \\
\left[ \mcQ | \mcQ \right]
    & = - e^{2\mu} \left( 2 \kappa + m^2 g_{(0)}^{\varphi\varphi} \right) + e^{2\psi-2\nu} \sin^2\theta  \left( r^2 \omega_{,r}^2 + \omega_{,\theta}^2  \right) - \frac{2}{r} \left( \psi_{,r} - 4 \nu_{,r} + 4 \mu_{,r} \right) \\
    & \qquad
    - \left( \psi_{,r}^2 - 4 \psi_{,r} \nu_{,r} - \nu_{,r}^2 - 4 \mu_{,r} \nu_{,r} + 4 \mu_{,r} ^{2} \right)
    - \frac{1}{r^2} \left(\psi_{,\theta}^2 - 4 \psi_{,\theta} \nu_{,\theta} - \nu_{,\theta}^2 - 4 \nu_{,\theta} \mu_{,\theta}  + 4 \mu_{,\theta} ^{2} \right) \nonumber \\
    & \qquad
    - \frac{1}{r^2} \left[ \left( 2 \psi_{,\theta}  - 4 \nu_{,\theta}  + \cot\theta \right) \cot\theta +5 \right] \nonumber\\
\left[ \mcQ | \mcQ_{,t} \right]
    & = - 2 im e^{2\mu-2\nu}\omega   \\
\left[ \mcQ | \mcQ_{,r} \right]
    & = \psi_{,r} + 3 \nu_{,r} + \frac{2}{r} \\
\left[ \mcQ | \mcQ_{,rr} \right]
    & = 1 \\
\left[ \mcQ | \mcQ_{,\theta} \right]
    & = \frac{1}{r^2} \left( \psi_{,\theta} +3 \nu_{,\theta} + \cot\theta \right) \\
\left[ \mcQ | \mcQ_{,\theta\theta} \right]
    & = \frac{1}{r^2} \\
\left[ \mcQ | \mcW \right]
    & =  -\frac{4}{r^2} \left[ \left( \psi_{,r} r + 1 \right) \left( \nu_{,\theta}- \mu_{,\theta}\right) + \left( \nu_{,r} r -\mu_{,r} r - 1\right) \left( \psi_{,\theta} +  \cot\theta \right) + \psi_{,r\theta} r \right] - 2 e^{2\psi-2\nu} \omega_{,r} \omega_{,\theta} r \sin^2\theta    \\
\left[ \mcQ | \mcY \right]
    & = 2 e^{2\psi} \sin\theta \left( \psi_{,\theta} - \mu_{,\theta} + \cot\theta \right) \omega_{,r} + 2 e^{2\psi} \sin\theta \left( \psi_{,r} - \mu_{,r} + \partial_r \right) \omega_{,\theta} \\
\left[ \mcW | \mcA \right]
    & = -im \left( {\frac {2 \left( \psi_{,r} r+1 \right) }{r^2 \sin\theta }} + e^{2\psi-2\nu}\omega\omega_{,r} r\sin\theta \right)   \\
\left[ \mcW | \mcA_{,t} \right]
    & = - e^{2\psi-2\nu} \omega_{,r} r \sin\theta  \\
\left[ \mcW | \mcB \right]
    & = -im \left( {\frac {2 \left( \psi_{,\theta} +\cot\theta \right) }{r^2 \sin\theta}}+  e^{2\psi-2\nu} \omega \omega_{,\theta} \sin\theta \right)   \\
\left[ \mcW | \mcB_{,t} \right]
    & = - e^{2\psi-2\nu} \omega_{,\theta} \sin\theta \\
\left[ \mcW | \mcH \right]
    & = e^{2\mu} \left(3 \kappa -2 \kappa_1 \right) - e^{2\psi-2\nu} \left( r^2 \omega_{,r}^2 + \omega_{,\theta}^2  \right) \sin^2\theta - 2 \left( \psi_{,r} + \frac{1}{r} \right) \nu_{,r} - \frac{2}{r^2}\left( \psi_{,\theta} + \cot\theta\right) \nu_{,\theta}   \\
\left[ \mcW | \mcK \right]
    & = e^{2\mu}\kappa + \frac{1}{2} e^{2\psi-2\nu} \left( 3 \omega_{,\theta}^2 -r^2 \omega_{,r}^2 \right) \sin^2\theta - 2 \left( \mu_{,r} + \frac{1}{r} \right) \nu_{,r} + 2 \left( \psi_{,r} + \frac{1}{r} \right)^2 \\
    & \qquad
    - \frac{2}{r^2} \left(\nu_{,\theta}^2 - \nu_{,\theta} \mu_{,\theta} +\nu_{,\theta\theta} \right) \nonumber \\
\left[ \mcW | \mcP \right]
    & = e^{2\mu} \left( 3 \kappa - 4 \kappa_1 \right) - e^{2\psi-2\nu} \omega_{,\theta}^2 \sin^2\theta - 2 \mu_{,r} \psi_{,r} - \frac{2}{r} \left( \psi_{,r} + \mu_{,r} \right) \\
    & \qquad
    + \frac{2}{r^2} \left[ \left(\psi_{,\theta} + \cot\theta \right) \mu_{,\theta} + \cot^2\theta - \psi_{,\theta\theta} \right]
    \nonumber \\
\left[ \mcW | \mcQ \right]
    & = - e^{2\psi-2\nu}  \omega_{,r} \omega_{,\theta} r \sin^2\theta + \frac{2}{r} \left [ \left( \psi_{,r} + \frac{1}{r} \right) \mu_{,\theta}+ \left( \mu_{,r} + \frac{1}{r} \right) \left( \psi_{,\theta} + \cot\theta \right) - \psi_{,r\theta} \right]  \\
\left[ \mcW | \mcW \right]
    & =  e^{2\mu} \left( \kappa - 2 \kappa_1 - m^2 g_{(0)}^{\varphi\varphi} \right) + e^{2\psi-2\nu} \left( r^2 \omega_{,r}^2 + \omega_{,\theta}^2  \right)  \sin^2\theta - 2 \left(\psi_{,r} + \frac{1}{r} \right)^2 - \frac{2}{r^2} \left(\psi_{,\theta} + \cot\theta \right)^2  \\
\left[ \mcW | \mcW_{,t} \right]
    & = - 2 im e^{2\mu-2\nu}\omega  \\
\left[ \mcW | \mcW_{,r} \right]
    & = \psi_{,r} + \nu_{,r} + \frac{2}{r} \\
\left[ \mcW | \mcW_{,rr} \right]
    & = 1 \\
\left[ \mcW | \mcW_{,\theta} \right]
    & = \frac{1}{r^2} \left( \psi_{,\theta} +\nu_{,\theta} +\cot\theta \right) \\
\left[ \mcW | \mcW_{,\theta\theta} \right]
    & = \frac{1}{r^2} \\
\left[ \mcW | \mcY \right]
    & =  -2\kappa_1 e^{2\psi+2\mu} \varpi r \sin\theta + e^{2\psi}\left( \psi_{,r} + 2 \nu_{,r} \right) \omega_{,r} r \sin\theta + \frac{ e^{2\psi}}{r}\left( \psi_{,\theta} + 2 \nu_{,\theta} \right) \omega_{,\theta} \sin\theta  \\
\left[ \mcW | \mcY_{,r} \right]
    & = e^{2\psi} \omega_{,r} r \sin\theta  \\
\left[ \mcW | \mcY_{,\theta} \right]
    & = \frac{e^{2\psi}}{r}\omega_{,\theta} \sin\theta \\
\left[ \mcW | Q_1 \right]
    & = -8 \pi \left( 2 \omega +\varpi \right) \varpi e^{2\psi+2\mu} r^2 \sin^2\theta  \\
\left[ \mcW | Q_2 \right]
    & = 16\pi e^{2\psi+2\mu} \varpi r \sin\theta  \\
\left[ \mcW | Q_6 \right]
    & = 8\pi e^{4\mu-2\nu}  \left( u^t \right)^{-2}  \\
\left[ \mcY | \mcA \right]
    & = -6 im e^{-2\nu}\omega \left( \psi_{,r} +\frac{1}{r} \right)  \\
\left[ \mcY | \mcA_{,t} \right]
    & = 2 e^{-2\nu} \nu_{,r}  \\
\left[ \mcY | \mcA_{,r} \right]
    & = -2 im e^{-2\nu} \omega \\
\left[ \mcY | \mcB \right]
    & = -\frac {2}{r} im e^{-2\nu} \omega \left( 3 \psi_{,\theta} + 2 \cot\theta \right) \\
\left[ \mcY | \mcB_{,t} \right]
    & = \frac { 2 }{r} e^{-2\nu} \nu_{,\theta} \\
\left[ \mcY | \mcB_{,\theta} \right]
    & = - \frac{2}{r} im e^{-2\nu} \omega \\
\left[ \mcY | \mcH_{,r} \right]
    & = - 2 e^{-2\nu} \omega_{,r} r\sin\theta  \\
\left[ \mcY | \mcH_{,\theta} \right]
    & = -\frac{2}{r} e^{-2\nu} \omega_{,\theta} \sin\theta \\
\left[ \mcY | \mcK \right]
    & = \frac{2}{r} e^{-2\nu} \sin\theta \left( 3  \psi_{,\theta} - \nu_{,\theta} + 3 \cot\theta + \partial_\theta \right) \omega_{,\theta}  \\
\left[ \mcY | \mcK_{,r} \right]
    & = -2 e^{-2\nu} \omega_{,r} r \sin\theta \\
\left[ \mcY | \mcL \right]
    & = -\frac {2 im \left( \psi_{,r} r+1 \right)  }{e^{2\psi+2\nu} r^2 \sin\theta} \\
\left[ \mcY | \mcM \right]
    & = -\frac {2 im \left( \psi_{,\theta} +\cot\theta \right)  }{e^{2\psi+2\nu} r^2 \sin\theta} \\
\left[ \mcY | \mcP \right]
    & =  -8 \kappa_1 e^{2\mu-2\nu}\varpi r\sin\theta - \frac{2}{r} e^{-2\nu} \sin\theta \left( 3 \psi_{,\theta} - \nu_{,\theta} + 3 \cot\theta +\partial_\theta \right) \omega_{,\theta}  \\
\left[ \mcY | \mcP_{,\theta} \right]
    & = - \frac{2}{r} e^{-2\nu} \omega_{,\theta} \sin\theta \\
\left[ \mcY | \mcQ \right]
    & = - e^{-2\nu} \sin\theta \left( 3  \psi_{,r} -  \nu_{,r}  + \partial_r + \frac{3}{r} \right) \omega_{,\theta} 
    - e^{-2\nu} \sin\theta \left( 3 \psi_{,\theta} - \nu_{,\theta} +3 \cot\theta  +\partial_\theta \right) \omega_{,r} \\
\left[ \mcY | \mcQ_{,r} \right]
    & = - e^{-2\nu} \omega_{,\theta} \sin\theta  \\
\left[ \mcY | \mcQ_{,\theta} \right]
    & = - e^{-2\nu} \omega_{,r} \sin\theta  \\
\left[ \mcY | \mcW \right]
    & = - 4 e^{2\mu-2\nu} \left( 2 \varpi \kappa_1 r\sin\theta + \frac {m^2\omega }{ e^{2\psi}r \sin\theta} \right) \\
\left[ \mcY | \mcW_{,r} \right]
    & = 2 e^{-2\nu} \omega_{,r} r \sin\theta \\
\left[ \mcY | \mcW_{,\theta} \right]
    & = \frac{2}{r} e^{-2\nu} \omega_{,\theta} \sin\theta \\
\left[ \mcY | \mcY \right]
    & =  e^{2\mu} \left( 8\pi(p-\epsilon) - 2 m^2 e^{-2\nu} \omega^2 -m^2g_{(0)}^{\varphi\varphi} \right) + e^{2\psi-2\nu}  \left( r^2 \omega_{,r}^2 + \omega_{,\theta}^2  \right) \sin^2\theta 
    + 4 \psi_{,r} \nu_{,r}  + \frac{1}{r} \left(\nu_{,r} - 3 \psi_{,r} \right)
    \\
    & \qquad  + \frac{1}{r^2} \left( 4 \psi_{,\theta} \nu_{,\theta} - 3 \psi_{,\theta} \cot\theta + \nu_{,\theta} \cot\theta - \cot^2\theta - 1 \right)   \nonumber \\
\left[ \mcY | \mcY_{,r} \right]
    & =  3 \psi_{,r} + 3 \nu_{,r} +\frac{2}{r}\\
\left[ \mcY | \mcY_{,rr} \right]
    & = 1 \\
\left[ \mcY | \mcY_{,\theta} \right]
    & = \frac{1}{r^2} \left( 3 \psi_{,\theta} + 3 \nu_{,\theta} + \cot\theta \right) \\
\left[ \mcY | \mcY_{,\theta\theta} \right]
    & = \frac{1}{r^2} \\
\left[ \mcY | Q_1 \right]
    & = -16\pi e^{2\mu}\omega r \sin\theta   \\
\left[ \mcY | Q_2 \right]
    & = 16\pi e^{2\mu}
\end{align}
\endgroup


\section{Equations Governing the Fluid Dynamics}
\label{app:fluid}

We present the perturbation equations for the neutron star fluid very similar to how we laid out the equations for the spacetime perturbation variables in Appendix~\ref{app:spacetime} in that we list all non-zero coefficients of the evolution equations. The crucial difference here is that the hydrodynamical equations, stemming from the perturbed law of conservation of energy-momentum, $\delta \left(\nabla^\mu T_{\mu\nu}\right) = 0$, contain only first derivatives of the fluid perturbations rather than taking the form of wave equations. For any evolved fluid perturbation, say, $Q_i$ the corresponding evolution equation can be written as 
\begin{align}
    \frac{\partial Q_i}{\partial t} = \sum_{\text{\Bicycle} \in \Sigma_2} \left[ Q_i | \text{\Bicycle} \right]  \text{\Bicycle} \quad\text{ for }i = 1, \ldots, 4,
\end{align}
where, again, the symbol $\left[ Q_i | \text{\Bicycle}\right]$ denotes the coefficient that appears in front of $\text{\Bicycle}$ in the evolution equation for $Q_i$, and $\Sigma_2$ is the set of the potentially appearing perturbation functions and their derivatives, i.e.,
\begin{align}
    \Sigma_2
        & := \bigcup_{i=1}^6 \left\{ Q_i, Q_{i,r}, Q_{i,\theta} \right\}
            \cup \bigcup_{\mcX \in \Xi} \left\{ \mcX, \mcX_{,t}, \mcX_{,r}, \mcX_{,\theta} \right\},
\end{align}
and $\Xi$ is defined as above. The non-zero coefficients of the fluid perturbation equations are (with the use of the definitions shown in Appendix~\ref{app:abbreviations}) as follows:
\begingroup
\allowdisplaybreaks
\begin{align}
\left[ Q_1 | \mcH \right]
    & = im \left( \kappa_2 \omega - \frac{1}{2} e^{-2\nu} \omega (\epsilon + p) \right) \\
\left[ Q_1 | \mcH_{,t} \right]
    & = \kappa_2 - \frac{1}{2} e^{-2\nu} \left( \epsilon + p \right) \\
\left[ Q_1 | \mcK \right]
    & = im \left( 2 \kappa_2 \Omega - \frac{1}{2} e^{-2\nu} \omega (\epsilon + p) \right) \\
\left[ Q_1 | \mcK_{,t} \right]
    & = 2\kappa_2 - \frac{1}{2} e^{-2\nu} \left( \epsilon + p \right) \\
\left[ Q_1 | \mcP \right]
    & = im \left( 2 \kappa_2 \Omega - \frac{1}{2} e^{-2\nu} \omega (\epsilon + p) \right) \\
\left[ Q_1 | \mcP_{,t} \right]
    & = 2\kappa_2 - \frac{1}{2} e^{-2\nu} \left( \epsilon + p \right) \\
\left[ Q_1 | \mcW \right]
    & = im \left( \kappa_2 \left(\omega + 2\varpi \right) + \frac{1}{2} e^{-2\nu} \omega (\epsilon + p) \right) \\
\left[ Q_1 | \mcW_{,t} \right]
    & = \kappa_2 + \frac{1}{2} e^{-2\nu} \left( \epsilon + p \right) \\
\left[ Q_1 | \mcY \right]
    & = - im \kappa_2 e^{2\psi} \varpi^2 r\sin\theta \\
\left[ Q_1 | Q_2 \right]
    & = -\frac{im}{r\sin\theta} \\
\left[ Q_1 | Q_3 \right]
    & = - \psi_{,r} - 3\nu_{,r} - 2\mu_{,r} - \frac{2}{r} - e^{2\psi-2\nu} \varpi \omega_{,r} r^2 \sin^2\theta \\
\left[ Q_1 | Q_{3,r} \right]
    & = -1 \\
\left[ Q_1 | Q_4 \right]
    & = - \frac{1}{r} \left( \psi_{,\theta} + 3\nu_{,\theta} + 2\mu_{,\theta} + \cot\theta \right)
        - e^{2\psi-2\nu} \varpi \omega_{,\theta} r \sin^2\theta \\
\left[ Q_1 | Q_{4,\theta}\right]
    & = - \frac{1}{r} \\
\left[ Q_2 | \mcH \right]
    & = im \left[ \kappa_2 \Omega^2
            - \frac{1}{2} \left( \epsilon + p \right)
                \left( \frac{1}{e^{2\psi} r^2 \sin^2\theta} + \frac{\omega^2}{e^{2\nu}} \right) \right] r\sin\theta \\
\left[ Q_2 | \mcH_{,t} \right]
    & = - \left[ \frac{1}{2} \left( \epsilon + p \right) \omega e^{-2\nu}
        - \kappa_2 \left( 2 \varpi + \omega \right)
        \right] r\sin\theta \\
\left[ Q_2 | \mcK \right]
    & = im \left[ 2\kappa_2 \Omega^2
            + \frac{1}{2} \left( \epsilon + p \right)
                \left( \frac{1}{e^{2\psi} r^2 \sin^2\theta} - \frac{\omega^2}{e^{2\nu}} \right) \right] r\sin\theta \\
\left[ Q_2 | \mcK_{,t} \right]
    & = - \left[ \frac{1}{2} \left( \epsilon + p \right) \omega e^{-2\nu}
        - 2 \kappa_2 \Omega
        \right] r \sin\theta \\
\left[ Q_2 | \mcP \right]
    & = im \left[ 2\kappa_2 \Omega^2
            + \frac{1}{2} \left( \epsilon + p \right)
                \left( \frac{1}{e^{2\psi} r^2 \sin^2\theta} - \frac{\omega^2}{e^{2\nu}} \right) \right] r \sin\theta \\
\left[ Q_2 | \mcP_{,t} \right]
    & = - \left[ \frac{1}{2} \left( \epsilon + p \right) \omega e^{-2\nu}
        - 2 \kappa_2 \Omega
        \right] r \sin\theta \\
\left[ Q_2 | \mcW \right]
    & = im \left[ \kappa_2 \Omega^2
            + \frac{1}{2} \left( \epsilon + p \right)
                \left( \frac{1}{e^{2\psi} r^2 \sin^2\theta} + \frac{\omega^2}{e^{2\nu}} \right) \right] r \sin\theta \\
\left[ Q_2 | \mcW_{,t} \right]
    & = \left[ \frac{1}{2} \left( \epsilon + p \right) \omega e^{-2\nu}
        + \kappa_2 \omega
        \right] r \sin\theta \\
\left[ Q_2 | \mcY \right]
    & = im \kappa_2 \omega \left[ e^{2\nu} - e^{2\psi} \varpi^2 r^2 \sin^2\theta \right] \\
\left[ Q_2 | \mcY_{,t} \right]
    & = \kappa_2 e^{2\nu} \\
\left[ Q_2 | Q_3 \right]
    & = - e^{2\psi-2\nu} \varpi \omega \omega_{,r} r^3 \sin^3\theta
    \nonumber\\
    & \qquad
        - \left[ \left( 3\varpi + \omega \right) \psi_{,r} r + \left( \varpi + 3\omega \right) \nu_{,r} r + 2 \left( \varpi + \omega \right) \mu_{,r} r - \omega_{,r} r + \left( 2 \varpi + \omega \right) 2 \right] \sin\theta \\
\left[ Q_2 | Q_{3,r} \right]
    & = - \Omega r \sin\theta \\
\left[ Q_2 | Q_4 \right]
    & = - e^{2\psi-2\nu} \varpi \omega \omega_{,\theta} r^2 \sin^3\theta
    \nonumber\\
    & \qquad
        - \left[ \left( 3\varpi + \omega \right) \psi_{,\theta}  + \left( \varpi + 3 \omega \right) \nu_{,\theta} + 2 \left( \varpi + \omega \right) \mu_{,\theta} - \omega_{,\theta} + \left( 3\varpi + \omega \right) \cot\theta \vphantom{\frac{1}{r}} \right] \sin\theta \\
\left[ Q_2 | Q_{4,\theta} \right]
    & = - \Omega \sin\theta \\
\left[ Q_2 | Q_5 \right]
    & = - \frac{im}{r\sin\theta} \\
\left[ Q_3 | \mcA \right]
    & = im \kappa_2 e^{2\psi-2\mu} \Omega \varpi r\sin\theta \\
\left[ Q_3 | \mcA_{,t} \right]
    & = \kappa_2 e^{2\psi-2\mu} \varpi r \sin\theta \\
\left[ Q_3 | \mcH \right]
    & = - \kappa_2 e^{2\nu-2\mu} \nu_{,r} - \kappa_2 \varpi e^{2\psi-2\mu} \left[ \varpi \left( \psi_{,r} r + 1 \right) - \omega_{,r} r \right] r \sin^2\theta\\
\left[ Q_3 | \mcH_{,r} \right]
    & = \frac{e^{-2\mu}}{2} \left( \epsilon + p \right) - \kappa_2 e^{2\nu-2\mu}  \\
\left[ Q_3 | \mcK \right]
    & = \kappa_2 e^{2\nu-2\mu} \nu_{,r} - \kappa_2 \varpi e^{2\psi-2\mu} \left[ \varpi \left( \psi_{,r} r + 1 \right) - \omega_{,r} r \right] r \sin^2\theta\\
\left[ Q_3 | \mcK_{,r} \right]
    & = \frac{e^{-2\mu}}{2} \left( \epsilon + p \right) \\
\left[ Q_3 | \mcL \right]
    & = -im \kappa_2 e^{-2\mu} \Omega \\
\left[ Q_3 | \mcL_{,t} \right]
    & = - \kappa_2 e^{-2\mu} \\
\left[ Q_3 | \mcP \right]
    & = \kappa_2 e^{2\nu-2\mu} \nu_{,r} - \kappa_2 \varpi e^{2\psi-2\mu} \left[ \varpi \left( \psi_{,r} r + 1 \right) - \omega_{,r} r \right] r \sin^2\theta\\
\left[ Q_3 | \mcP_{,r} \right]
    & = \frac{e^{-2\mu}}{2} \left( \epsilon + p \right) \\
\left[ Q_3 | \mcW \right]
    & = \kappa_2 e^{2\nu-2\mu} \nu_{,r} + \kappa_2 \varpi e^{2\psi-2\mu} \left[ \varpi \left( \psi_{,r} r + 1 \right) - \omega_{,r} r \right] r \sin^2\theta\\
\left[ Q_3 | \mcW_{,r} \right]
    & = - \frac{e^{-2\mu}}{2} \left( \epsilon + p \right) + \kappa_2 e^{2\nu-2\mu} \\
\left[ Q_3 | \mcY \right]
    & = \kappa_2 e^{2\psi+2\nu-2\mu} \left[ \omega_{,r} r - \varpi \left( 2\psi_{,r} r + 2\nu_{,r} r + 1 \right) \right] \sin\theta \\
\left[ Q_3 | \mcY_{,r} \right]
    & = - \kappa_2 e^{2\psi+2\nu-2\mu} \varpi r \sin\theta \\
\left[ Q_3 | Q_1 \right]
    & = - e^{2\nu-2\mu} \nu_{,r} + e^{2\psi-2\mu} r \sin^2\theta
        \left[ \omega_{,r} \omega r - \left( \psi_{,r} r + 1 \right ) \left( 2 \omega + \varpi \right) \varpi \right] \\
\left[ Q_3 | Q_2 \right]
    & = e^{2\psi-2\mu} \sin\theta \left[ 2\varpi \left(\psi_{,r} r + 1 \right) - \omega_{,r} r \right] \\
\left[ Q_3 | Q_3 \right]
    & = -im \Omega \\
\left[ Q_3 | Q_6 \right]
    & = - \left[ \nu_{,r} + 2\mu_{,r} + e^{2\psi-2\nu} \varpi^2 r \sin^2\theta \left(\psi_{,r} r + 1 \right) \right] \\
\left[ Q_3 | Q_{6,r} \right]
    & = -1 \\
\left[ Q_4 | \mcB \right]
    & = im \kappa_2 e^{2\psi-2\mu} \varpi \Omega r \sin\theta \\
\left[ Q_4 | \mcB_{,t} \right]
    & = \kappa_2 e^{2\psi-2\mu} \varpi r \sin\theta \\
\left[ Q_4 | \mcH \right]
    & = - \frac{\kappa_2 e^{2\nu-2\mu} \nu_{,\theta}}{r}
        - \kappa_2 \varpi e^{2\psi-2\mu} r \sin^2\theta \left[ \varpi \left( \psi_{,\theta} + \cot\theta \right) - \omega_{,\theta} \right] \\
\left[ Q_4 | \mcH_{,\theta} \right]
    & = \frac{e^{-2\mu}}{2r} \left( \epsilon + p \right) - \frac{\kappa_2 e^{2\nu-2\mu}}{r} \\
\left[ Q_4 | \mcK \right]
    & = \frac{\kappa_2 e^{2\nu-2\mu} \nu_{,\theta}}{r}
        - \kappa_2 \varpi e^{2\psi-2\mu} r \sin^2\theta \left[ \varpi \left( \psi_{,\theta} + \cot\theta \right) - \omega_{,\theta} \right] \\
\left[ Q_4 | \mcK_{,\theta} \right]
    & = \frac{e^{-2\mu}}{2r} \left( \epsilon + p \right) \\
\left[ Q_4 | \mcM \right]
    & = - im \kappa_2 e^{-2\mu} \Omega \\
\left[ Q_4 | \mcM_{,t} \right]
    & = - \kappa_2 e^{-2\mu} \\
\left[ Q_4 | \mcP \right]
    & = \frac{\kappa_2 e^{2\nu-2\mu} \nu_{,\theta}}{r}
        - \kappa_2 \varpi e^{2\psi-2\mu} r \sin^2\theta \left[ \varpi \left( \psi_{,\theta} + \cot\theta \right) - \omega_{,\theta} \right] \\
\left[ Q_4 | \mcP_{,\theta} \right]
    & = \frac{e^{-2\mu}}{2r} \left( \epsilon + p \right) \\
\left[ Q_4 | \mcW \right]
    & = \frac{\kappa_2 e^{2\nu-2\mu} \nu_{,\theta}}{r}
        + \kappa_2 \varpi e^{2\psi-2\mu} r \sin^2\theta \left[ \varpi \left( \psi_{,\theta} + \cot\theta \right) - \omega_{,\theta} \right] \\
\left[ Q_4 | \mcW_{,\theta} \right]
    & = - \frac{e^{-2\mu}}{2r} \left( \epsilon + p \right) + \frac{\kappa_2 e^{2\nu-2\mu}}{r} \\
\left[ Q_4 | \mcY \right]
    & = - \kappa_2 e^{2\psi+2\nu-2\mu} \sin\theta \left[ \varpi \left( 2\psi_{,\theta} + 2\nu_{,\theta} + \cot\theta \right) - \omega_{,\theta} \right] \\
\left[ Q_4 | \mcY_{,\theta} \right]
    & = - \kappa_2 e^{2\psi+2\nu-2\mu} \varpi \sin\theta \\
\left[ Q_4 | Q_1 \right]
    & = - \frac{e^{2\nu-2\mu} \nu_{,\theta}}{r} - e^{2\psi-2\mu} r \sin^2\theta \left[ \left( \psi_{,\theta} + \cot\theta \right) \left( 2 \omega + \varpi \right) \varpi - \omega_{,\theta} \omega \right] \\
\left[ Q_4 | Q_2 \right]
    & = e^{2\psi-2\mu} \sin\theta \left( 2\varpi \left( \psi_{,\theta} + \cot\theta \right) - \omega_{,\theta} \right) \\
\left[ Q_4 | Q_4 \right]
    & = - im \Omega \\
\left[ Q_4 | Q_6 \right]
    & = - \frac{1}{r} \left( 2\mu_{,\theta} + \nu_{,\theta} \right) - e^{2\psi-2\nu} \varpi^2 r \sin^2\theta \left( \psi_{,\theta} + \cot\theta \right) \ \\
\left[ Q_4 | Q_{6,\theta} \right]
    & = - \frac{1}{r}
\end{align}
\endgroup

\end{widetext}

%

\end{document}